\documentclass[11pt]{article}
\usepackage[utf8]{inputenc}
\usepackage{amsmath,amssymb}
\usepackage[margin = 1in]{geometry}
\usepackage{algorithm,algorithmicx,algpseudocode}
\usepackage{natbib}
\usepackage{graphicx}
\usepackage{setspace}
\usepackage{booktabs}
\usepackage{subcaption} 
\usepackage{float} 
\usepackage{tabularx}
\usepackage{adjustbox}

\usepackage{hyperref} 
\hypersetup{	colorlinks=true,	linkcolor=blue,	filecolor=magenta,	urlcolor=blue,	allcolors=blue,}
\def\spacingset#1{\renewcommand{\baselinestretch}%
{#1}\small\normalsize} \spacingset{1}

\usepackage{tikz}

\newcommand{\cm}[1]{\ignorespaces}
\usepackage{xcolor}
\definecolor{mypink}{RGB}{219, 48, 122}

\newcommand\numberthis{\addtocounter{equation}{1}\tag{\theequation}}

\def\diag{\text{diag}}
\def\rest{\text{rest}}

\newcommand{\sbr}[1]{\left( #1 \right) }
\newcommand{\br}[1]{\left\{ #1 \right\} }
\newcommand{\mbr}[1]{\left[ #1 \right] }

\def\bfdelta{\mathbf \delta}

\def\indi{\mathbf 1}

\def\bfY{\mathbf Y}

\def\bE{\mathbb E}
\def\bP{\mathbb P}
\def\dd{\text{d}}

\def\bfbeta{\boldsymbol \beta}
\def\bftheta{\boldsymbol \theta}
\def\bfepsilon{\boldsymbol \epsilon}
\def\bfdelta{\boldsymbol \delta}

\def\cB{\mathcal B}
\def\cN{\mathcal N}

\def\diag{\mathrm{diag}}

\def\R{\mathbb R}

\def\cV{\mathcal{V}}
\def\cM{\mathcal{M}}

\newif\ifhidetext
\hidetextfalse 

\newcommand{\hidetext}[1]{%
  \ifhidetext
  \else
    #1
  \fi
}

\title{Scalable Bayesian Image-on-Scalar Regression for Population-Scale Neuroimaging Data Analysis}
\author{ \hidetext{Yuliang Xu$^1$\thanks{Most of this work was completed during Yuliang Xu's PhD training in the Department of Biostatistics at the University of Michigan.}, Timothy D. Johnson$^2$, Thomas E. Nichols$^3$, and\\ Jian Kang$^2$\thanks{To whom correspondence should be addressed: jiankang@umich.edu}\\[4mm]
    Department of Statistics, University of Chicago$^1$\\
    Department of Biostatistics, University of Michigan$^2$\\
    Big Data Institute, University of Oxford$^3$} }

\begin{document}

\maketitle

\begin{abstract}

Bayesian Image-on-Scalar Regression (ISR) provides flexible, uncertainty-aware neuroimaging analysis. However, applying ISR to large-scale datasets such as the UK Biobank is challenging due to intensive computational demands and the need to handle subject-specific brain masks rather than a common mask. We propose a novel Bayesian ISR model that scales efficiently while accommodating these inconsistent masks. Our method leverages Gaussian process priors with salience area indicators and introduces a scalable posterior computation algorithm using stochastic gradient Langevin dynamics combined with memory mapping. This approach achieves linear scaling with subsample size and constrains memory usage to the batch size, facilitating direct spatial posterior inferences on brain activation regions. Simulation studies and analysis of UK Biobank task fMRI data (38,639 subjects; over 120,000 voxels per image) demonstrate a 4- to 11-fold speed increase and an 8–18\% enhancement in statistical power compared to traditional Gibbs sampling with zero-imputation. Our analysis reveals a subregion of the amygdala where emotion-related brain activation decreases by approximately 58\% between ages 50 and 60.

\end{abstract}
\noindent%
\vfill

\newpage
\spacingset{1.2}
\section{Introduction}

Magnetic Resonance Imaging (MRI) is a non-invasive technique renowned for its comprehensive insight into the brain's structure and function. Functional MRI (fMRI) is an imaging modality that detects neuronal activity via fluctuations in the blood oxygen level dependent (BOLD) signal, providing insights into brain activity \citep{Lindquist2008-ej}. Task-based fMRI can be used to identify regions where individual traits (e.g. cognitive ability) associate with brain function. We aim to map the influence of a single trait on activity over the whole brain. This represents a classical problem in imaging statistics, where the outcome is an image, and the predictors are multiple scalar variables, commonly known as Image-on-Scalar regression (ISR). As an example, using UK Biobank data \citep{sudlow2015uk}, we focus on examining the influence of age on brain activity across the whole brain using task fMRI data. In this scenario, the task fMRI data is the outcome image, while age is the scalar predictor variable. 

The analysis of large-scale brain fMRI data presents significant challenges including low signal-to-noise ratio and complex anatomical brain structure \citep{Lindquist2008-ej,smith2018statistical}. The advent of large-scale neuroimaging studies, such as the UK Biobank and Adolescent Brain Cognitive Development (ABCD) study, has introduced new computational challenges for traditional statistical tools in analyzing large-scale fMRI data. These challenges arise from the substantial size of the datasets, as well as the scalability of posterior computation algorithms and the difficulties encountered in achieving convergence in high-dimensional settings.  

A specific challenge associated with the UK Biobank fMRI data is the presence of individual-specific brain masks. The brain typically occupies less than half of the cuboid image volume, and a brain mask is used to identify which voxels constitute brain parenchyma (functional brain tissue) and should be included in the analysis. Even after registration of brain images to standard space, each subject's fMRI data can have a unique brain mask, and this is the case in the UK Biobank task fMRI data.

In this paper, we seek to address these challenges by presenting a Bayesian hierarchical model for Image-on-Scalar regression (ISR) that incorporates a sparse and spatially correlated prior on the exposure coefficient. Furthermore, we propose an efficient algorithm utilizing Stochastic Gradient Langevin Dynamics \citep[SGLD]{Welling_undated-zi} to ensure scalability. To accommodate the varying individual-specific masks, we employ imputation techniques, enabling us to analyze a wide spatial mask across all individuals.

\subsection{UK Biobank Data}
The use of imaging biomarkers in clinical diagnostics and disease prognostics has been historically hindered by the lack of imaging data collected before disease onset. The UK Biobank is collecting longitudinal data on one million UK residents of which a sub-sample of one hundred thousand are being imaged longitudinally \citep{Miller2016-tq}. The UK Biobank data provides multimodal brain imaging data including structural, diffusion, and functional MRI data. We focus on the task fMRI data, an emotion task where participants are asked to identify faces with negative emotions using shape identification as the baseline task. The objective of this emotional task is to actively involve cognitive functions ranging from sensory and motor areas to regions responsible for processing emotions. Recent studies using the UK Biobank have utilized multimodal data to predict brain age \citep{cole2020multimodality}, have analyzed the association between resting state connectivity data, education level, and household income \citep{shen2018resting}, and have applied deep learning to sex classification in resting state and task fMRI connectomes \citep{leming2021deep}; amongst others \citep{elliott2018genome,littlejohns2020uk}.

\subsection{Traditional and Recent Practices in ISR}
The most common practice for Image-on-Scalar regression is the mass univariate analysis approach (MUA) \citep{Groppe2011-gv}. Although MUA is computationally efficient, easy to implement, and has well-developed multiple comparison correction methods to control false discovery rates, MUA ignores spatial structure and tends to have low statistical power when the data has a low signal-to-noise ratio. To account for spatial dependence,
as discussed in \cite{Morris2015-aj}, a common approach is to use a low-rank approximation. Built on the principle components idea to utilize spatial correlation, \cite{ramsay2005fitting} and \cite{Reiss2010-ql} proposed a penalized regression method using basis expansion to reduce the dimension of the functional outcome. \cite{Zhu2013-mk} propose another model that uses local polynomial and kernel methods for estimating the spatially varying coefficients with discontinuity jumps.
\cite{Yu2021-kz} and \cite{Li2021-by} use bivariate spline functions to estimate the spatial functional estimator supported on the 2-dimensional space. \cite{Zhang2021-ui} propose a quantile regression method that can be applied to high-dimensional outcomes. A common issue with all the aforementioned frequentist methods is that it can be difficult to make inferences on the active area selection based on these penalized low-rank models. \cite{Zhang2020-dh} develop an efficient deep neural network approach to estimate the spatially varying parameters of very high-dimension with complex structures, but they only focus on point estimation rather than statistical inference.
\cite{zeng2022bayesian} propose a Bayesian model with a prior composed of a latent Gaussian variable and a binary selection variable to account for both sparsity and spatial correlation. However, the dense covariance matrix can require large memory and computational power when the outcome is very high-dimensional. To address these limitations, we propose a scalable Bayesian Image-on-Scalar regression model where the functional coefficient is assigned a sparse and spatially correlated prior so that we can make inferences on the activation areas directly from the posterior inclusion probability (PIP).

\subsection{Subject-specific Masks in Brain Imaging}\label{subsec:mask}
The brain mask used for a multisubject fMRI analysis is typically the intersection of the individual-specific masks, only analyzing voxels where all subjects have data. However, in the UK Biobank, an intersection mask of 38,639 subjects reduces the analysis volume by 71\% relative to the average subject mask volume. Some authors will attempt to minimize this effect by identifying subjects with particularly small masks, but this is a laborious process and discards data. For a particular subject, all voxels that lie in the set difference of the union mask and that subject's mask results in missing values for said subject. Different from the missing data literature where the missing values exist in reality but are unobserved, the missing values here are due to the differently aligned individual masks. The goal is to augment individual image data to a common brain mask through imputation methods. Many of the voxels with missing data occur around the edge of the union mask \citep{Mulugeta2017-lr}. Historically, researchers usually do not account for missing data, however, for PET data, missing values are imputed using a \textit{soft mean} \citep{hammers2007upregulation} derived using available data at each voxel. A concurrent work \citep{lu2025new} introduced an imputation approach for imaging data based on the conditional distribution of missing voxels given the observed voxels. While this represents a useful contribution, their framework assumes a common set of missing voxels across all subjects, whereas in our setting the pattern of missingness varies across individuals. In addition, their method has so far been demonstrated on datasets of more modest scale (around 1000 voxels), which differs from the much larger-scale applications we consider. \label{sent:imputation}

 In volumetric fMRI data, individual brain masks can be slightly different from one another, due to some subject's brain falling outside of the field-of-view truncation, residual variation in inter-subject brain shape not accounted for by atlas registration, and susceptibility-induced signal loss. In particular,  voxels above the nasal cavity and above the ear canals suffer from signal loss and are classified as non-brain tissue in a subject-specific manner.
 
 All brain image analyses require a brain mask to avoid wasted computation and uninterpretable results on non-brain voxels.
 For mass-univariate fMRI analyses, some authors created custom software \citep{szaflarski2006longitudinal, maullin2022blmm} or used mixed effect models with longitudinal fMRI data on each voxel \citep{szaflarski201210} to account for subject-specific masks, or explicitly accounted for missing data with multiple imputation \citep{vaden2012multiple}. Among the major fMRI software packages, however, neither FSL nor SPM can account for subject-specific masks, and only AFNI's 3dMEMA \citep{chen2012fmri} (simple group analysis) or 3dLME \citep{chen2013linear} (general mixed effects)  accounts for subject-specific masks. 
 However, all of these methods are mass-univariate. To the best of our knowledge, our work is the first spatial Bayesian method to account for varying missing data patterns over the brain. By using individual masks with imputation, we make the most use of all collected data.

\subsection{Scalable Posterior Algorithms}
The scalable posterior algorithm we use is based on the stochastic gradient Langevin dynamics (SGLD) algorithm \citep{Welling_undated-zi}. The SGLD algorithm is effective at handling large-scale data as it approximates the posterior gradient using subsamples of the data. Variations on the SGLD algorithm for scalable posterior sampling have been proposed. \cite{Wu2022-pi} proposed a Metropolis-Hasting algorithm using mini-batches where the proposal and acceptance probability are both approximated by the current mini-batch. \cite{Kim2022-dg} proposed an adaptive SGLD algorithm (Adam SGLD) that sets a preconditioner for SGLD and allows the gradient at different directions to update with different step sizes. Aside from these MCMC algorithms, variational Bayesian inference \citep{jaakkola1999variational} is another popular option for approximating the posterior mean in high-dimensional settings. There has been increasing use of variational inference for high-dimensional posteriors such as imaging data analysis \citep{kaden2008variational, kulkarni2022mixed}, and various scalable extensions of variational inference \citep{Hoffman2013-qt, pmlr-v33-ranganath14, blaiotta2016variational}. Although these variational methods are computationally efficient, our goal, however, is to obtain the entire MCMC sample that provides uncertainty quantification for the activation areas of interest. Hence, we settled on the SGLD-type algorithm for its scalability.

The main contributions of our proposed method are
\begin{enumerate}
    \item to provide an efficient posterior computation algorithm for Bayesian Image-on-Scalar regression, scalable to large sample size and high-resolution image;
    \item to introduce the individual-specific brain masks and expand the analysis region from an intersection mask of all individuals to an inclusive mask using imputation.
\end{enumerate}
In particular, our method uses batch updates and memory-mapping techniques to analyze large sample imaging data that is too big to fit into random access memory on many computers. In addition, we provide an imputation-based method that allows us to handle individual-specific masks and makes full use of the observed data.

This paper is organized as follows. Section \ref{sec:model} introduces our Bayesian Image-on-Scalar model; Section \ref{sec:computation} details the algorithm and computational aspects;  Section \ref{sec:RDA} applies the method to the UK Biobank imaging data and presents sensitivity analyses; and Section \ref{sec:discussion} summarizes our contributions and discusses our findings. We defer the simulation results to Supplementary Section~\ref{sec:simulation}. Additional real data analysis results are provided in the Supplementary Materials. Our implementation is provided as an R package, SBIOS, and is publicly available on Github \footnote{ See the SBIOS R package on Github page \hidetext{\url{https://github.com/yuliangxu/SBIOS}} or in the Supplementary}.

\section{Model}\label{sec:model}

Let $\mathcal{B}\subset \mathbb{R}^3$ denote the entire brain region. Let $\br{s_j}_{j=1}^p \subset \cB$ be a set of fixed grid points in $\mathcal{B}$, on which we observe brain image intensity values. For individual $i~ (i=1,\dots,n)$, let $Y_i(s_j)$ be the image intensity at voxel $s_j$. To incorporate the individual-specific masks,  let $\cV_i$ denote the set of locations where the image intensity is observed for individual $i$, i.e., for any $s_j\in \cV_i$, $Y_i(s_j)$ is not missing. For any $i = 1,\dots, n$, $\cV_i \subset \br{s_1,\dots, s_p}.$
Let $X_i$ be the primary covariate of interest, $Z_{ik}$ be the $k$-th confounding covariate for $k=1,\dots,q$.
We propose an Image-on-Scalar regression model. For individual $i$ and any $s_j\in \cV_i$,
\begin{equation}
    Y_i(s_j) = X_i \beta(s_j)\delta(s_j) + \sum_{k=1}^m\gamma_k(s_j)Z_{ik} + \eta_i(s_j) + \epsilon_i(s_j), \quad \epsilon_i(s_j) \sim \mathrm{N}(0,\sigma_Y^2). \label{eq:model}
\end{equation}

The spatially varying parameter $\beta(s)$ estimates the magnitude in the image intensity that can be explained by the predictor $X$, and the binary selection indicator $\delta(s)$ follows a Bernoulli prior with selection probability $p(s)$. In practice, we set $p(s)=0.5$ for any $s\in \mathcal{B}$ as the prior for $\delta(s)$. The selection variable $\delta(s)$ determines the active voxels in the brain associated with the predictor $X$ and $\beta(s)\delta(s)$ is of main interest.  The spatially varying parameter $\gamma_k(s)$ is the coefficient for the $k$-th confounder $Z_k$, and $\eta_i(s)$ accounts for individual level spatially correlated noise. By introducing the individual effect $\eta_i$ as a parameter, we are separating spatially correlated noise from spatially independent noise $\epsilon_i$ and we can safely assume a completely independent noise term $\epsilon_i(s)$ across all locations $s_j$, hence avoiding large-scale covariance matrix computations in the noise term. This is similar to the correlated noise model in \cite{Zhu2013-mk}.

For model \eqref{eq:model}, we specify the following priors:
\begin{align}
    \delta(s) &\sim \mathrm{Ber}\{p(s)\}, \quad \mbox{ for any } s \in \mathcal{B} \label{eq:delta_prior}\\
    \beta(s) &\sim \mathcal{GP}(0,\sigma_\beta^2\kappa), \quad \mbox{ for any } s \in \mathcal{B} \label{eq:beta_prior}\\
    \gamma_k(s) &\sim \mathcal{GP}(0,\sigma_\gamma^2\kappa), \quad \mbox{ for any } s \in \mathcal{B} , \quad k=1,\dots,m \label{eq:gamma_prior} \\
    \eta_i(s) &\sim \mathcal{GP}(0,\sigma_\eta^2\kappa),\quad \mbox{ for any } s \in \mathcal{B} \label{eq:eta_prior} 
\end{align}
The spatially-varying functional coefficients $\beta(s),\gamma_k(s),\eta_i(s)$ are assumed to have Gaussian Process (GP) priors with mean 0 and kernel function $\sigma^2\kappa(\cdot,\cdot)$, where $\sigma^2$ can be different for each functional parameter. The priors in \eqref{eq:delta_prior} to \eqref{eq:eta_prior} are mutually independent on the prior level. Popular choices of the kernel function $\kappa$ include the exponential square kernel and the Mat\'ern kernel \eqref{eq:matern}. 
\begin{align}\label{eq:matern}
    \kappa(s',s;\nu,\rho) = C_\nu(\|s'-s\|_2^2/\rho), ~ C_\nu(d):= \frac{2^{1-\nu}}{\Gamma(\nu)}\left( \sqrt{2\nu} d \right)^\nu K_\nu(\sqrt{2\nu }d),
\end{align}
where $K_\nu$ is a modified Bessel function of the second kind \citep{Rasmussen2005-uh}.
We use the Mat\'ern kernel in both simulation studies and real data analysis as it offers flexible choices of the kernel parameters. Here, all GPs are assumed to have the same kernel function $\kappa$ for computational efficiency. In the UK Biobank data analysis, $\kappa$ is chosen to reflect the outcome image $Y_i(s)$ correlation structure in a data-adaptive way, as discussed in Section \ref{sec:RDA}.

Model \eqref{eq:model} can be extended to include a selection variable $\delta$ for multiple terms. 
\begin{equation}
    Y_i(s_j) = \sum_{h=1}^H X_{i,} \beta_{h}(s_j)\delta_{h}(s_j) + \sum_{k=1}^m\gamma_k(s_j)Z_{ik} + \eta_i(s_j) + \epsilon_i(s_j), \quad \epsilon_i(s_j) \sim \mathrm{N}(0,\sigma_Y^2). \label{eq:model_extend}
\end{equation}
In Section~\ref{supp_sec:RDA_delta}, we provide an analysis on the UKB data where a common $\delta(s)$ is applied to the main effect and the interaction effect. For the rest of this paper, we follow model \eqref{eq:model} and only view model \eqref{eq:model_extend} as an extended framework.

\section{Posterior Computation}\label{sec:computation}
\subsection{Posterior Sampling with Gaussian Process Priors}\label{sec:computation_GP}
The 3D task fMRI image data is divided into $R$ regions using the Harvard-Oxford cortical and subcortical structural atlases \citep{Desikan2006-wt}. Between-region independence is assumed when constructing the Gaussian kernel for $\beta$, $\gamma_k$ and $\eta_i$. Instead of assuming a whole brain correlation structure for the GP  priors, a block diagonal covariance structure is computationally efficient and allows us to capture more detailed information within each region, especially regions of smaller size and complex spatial structures.  

To sample from the posterior of the GPs, we use a basis decomposition approach. By Mercer's theorem \citep{Rasmussen2005-uh},
for any $\beta(s)\sim \mathcal{GP}(0,\sigma_\beta^2\kappa)$, we can use a basis decomposition,
\[\beta(s) = \sum_{l=1}^\infty \theta_{\beta,l} \psi_l(s),\quad \theta_{\beta,l}\sim \mathcal{N}(0,\sigma_\beta^2\lambda_l),\]
where $\lambda_l$ is the $l$-th eigenvalue, and $\psi_l$ is the $l$-th eigenfunction (see Section 4.2 in \cite{Rasmussen2005-uh}). The eigenvalues in this expansion satisfy $\sum_{l=1}^\infty \lambda_l <\infty$, and the eigenfunctions form an orthonormal basis in $L^2(\mathcal{B})$, i.e. $\int_{s\in \mathcal{B}} \psi_l(s) \psi_{l'}(s)d s = I(l=l')$, where $I(\cdot)$ is an indicator function taking value 1 if the expression inside the bracket is true. Hence, we use the coefficient space of $\theta_{\beta,l}$ rather than $\beta(s)$. Similarly, we expand $\gamma_k(s) = \sum_{l=1}^\infty \theta_{\gamma,k,l} \psi_l(s)$ and $\eta_i(s) = \sum_{l=1}^\infty \theta_{\eta,i,l}\psi_l(s)$, where $\theta_{\gamma,k,l}$ and $\theta_{\eta,i,l}$ are the basis coefficients for $\gamma_k$ and $\eta_i$ respectively. 
In practice, only a finite number $l=1,\dots,L$ of $\{\theta_{\beta,l}\}_{l=1}^\infty$ are used (corresponding to the $L$ largest eigenvalues), and $\beta(s)$ is approximated by $\sum_{l=1}^L\theta_{\beta,l}\psi_l(s)$. This basis decomposition approach is applied for all three GP priors \eqref{eq:beta_prior} - \eqref{eq:eta_prior}.

With the basis expansion coefficient, we can sample from the $L$-dimensional space to model the $p$-dimensional image data.
We also partition the brain into regions to further speed up computation and assume a region-independence structure for the spatially varying parameters $\beta,\gamma_k$, and $\eta_i$. Assume there are $r = 1,\dots, R$ regions that form a partition of the mask $\cB$, denoted as $\cB_1,\dots, \cB_R$. For the three GP priors \eqref{eq:beta_prior} - \eqref{eq:eta_prior}, we assume that the kernel function $\kappa(s_j, s_k) = 0$ for any $s_j\in \cB_r, s_k\in\cB_{r'}, r\neq r'$, and the prior covariance matrix on the fixed grid has a block diagonal structure. 

For the $r$-th region, let $p_r$ be the number of voxels in $\cB_r$ and let $Q_r = \sbr{\psi_l(s_{r,j})}_{l=1,j=1}^{L_r,p_r} \in \R^{ p_r\times L_r}$ be the matrix with the $(j,l)$-th component $\psi_l(s_{r,j})$ where $\br{s_{r,j}}_{j=1}^{p_r}$ forms the fixed grid in $\cB_r$. 
With the region partition, the GP priors on the $r$-th region can be reexpressed as $\bfbeta_r = \sbr{\beta(s_{r,1}),\dots,\beta(s_{r,p_r})}^T \approx Q_r\bftheta_{\beta,r}$, where $\bftheta_{\beta,r}\sim \cN\sbr{0,\sigma_\beta^2 D_r}$ and $D_r$ is a diagonal matrix with diagonal $(\lambda_{r,1},\dots,\lambda_{r, L_r})\in \R^{L_r}$. Note that $Q_r$ is not necessarily orthonormal as the finite approximation of the eigenfunctions, hence in practice we apply QR decomposition on the matrix formed by eigenfunctions, and take the Q-matrix as the final approximation of $Q_r$ to guarantee orthonormality.

To present the working model with region partitions, denote $\bfY^*_{i,r}=Q_r^T \bfY_{i,r} \in \R^{L_r}$ as the low-dimensional mapping of the $i$th image on the $r$-th region where $\bfY_{i,r} = \br{Y_i(s_j)}_{s_j\in \cB_r}\in \R^{p_r}$, and $\bfepsilon^*_{i,r}= Q_r^T \bfepsilon_{i,r}$, $\bfepsilon_{i,r} = \br{\epsilon_{i}(s_j)}_{s_j\in\cB_r}\in \R^{p_r}$. Let $\diag\br{x}$ be the diagonal matrix with diagonal $x$. Let $\bfdelta_r = \br{\delta(s_j)}_{s_j\in\cB_r}\in \R^{p_r}$. After basis decomposition
\begin{equation}\label{eq:working_model}
    \bfY_{i,r}^* = Q_r^T X_i\diag\{\bfdelta_r\} Q_r\bftheta_{\beta,r} + \sum_{k=1}^{m}\theta_{\gamma,k,r} Z_{i,k}+\bftheta_{\eta,i,r} + \bfepsilon^*_{i,r}
\end{equation}
with the prior specification $\bftheta_{\beta,r} \sim  N(0,\sigma_\beta^2 D_r)$, $\bftheta_{\gamma,k,r}\sim  N(0,\sigma_\gamma^2 D_r)$, $\bftheta_{\eta,i,r}\sim  N(0,\sigma_\eta^2 D_r)$, and the low-dimensional noise $\bfepsilon^*_{i,r}\sim N(0,\sigma_Y^2 I_{L,r})$. The working model \eqref{eq:working_model} based on regionally independent kernels performs a whole brain analysis since $\sigma_\beta^2, \sigma_\gamma^2, \sigma_\eta^2$, and $\sigma_Y^2$ are estimated globally across regions. This is also the first step towards reducing memory cost by using a low-dimensional approximation.
The finite cutoff $L$ is chosen to reflect the flexibility of the true functional parameters: fewer bases are required to approximate the smooth function $\beta(s)$. In our real data analysis, $L$ is determined by extracting the eigenvalues of the covariance kernel matrix. For one brain region, first compute the $p\times p$ dimensional covariance matrix with appropriately tuned covariance parameters, get the eigenvalues of such covariance matrix, and choose the cutoff such that the summation $\sum_{l=1}^L\lambda_l$ is over $90\%$ of $\sum_{l=1}^p\lambda_l$. Details for choosing the covariance parameters in \eqref{eq:matern} and sensitivity analysis on the $90\%$ cutoff are discussed in Sections \ref{sec:RDA} and \ref{sec:RDA_sensi}.

\subsection{Scalable Algorithm for Large Dataset}\label{subsec:scal_algo}

Inspired by \cite{Welling_undated-zi} and \cite{Wu2022-pi}, we propose Algorithm \ref{algo1} based on SGLD to sample the posteriors from a large data set. The SGLD algorithm outperforms Gibbs sampling (GS) in three ways: 1) under the assumption that all individuals are independently distributed according to the proposed model, the SGLD algorithm allows us to compute the log-likelihood on a small subset of data, making it computational efficient compared to GS; 2) the SGLD algorithm adds a small amount of noise to the gradient of the log posterior density, making it more effective in exploring the posterior parameter space; and 3) as the step size decreases to 0, the SGLD algorithm provides a smooth transition from the stochastic optimization stage to the posterior sampling stage.

We apply the SGLD algorithm on $\bftheta_\beta$. Denote $\bftheta_\beta^{(t)}$ as the value at the $t$-th iteration. Let $\tau_t$ be the step size at the $t$-th iteration, let $\pi(\theta_{\beta,l})$ denote the prior, and let $\pi_{i\in \mathcal{I}}\left(Y_i\mid X_i,Z_i, \theta\right):=\prod_{i\in \mathcal{I}}\pi\left(Y_i\mid X_i,Z_i, \theta\right)$ be the likelihood for a subsample $\mathcal{I}$, where $\theta$ is the collection of all parameters. Denote by $n_s$ the number of subjects in the subsample $\mathcal{I}$. Let $\nabla f(x)$ be the gradient of $f(x)$. At the $t$-th iteration, the $l$-th component in $\bftheta_\beta$ is updated as
\begin{equation}\label{eq:sgld}
    \theta_{\beta,l}^{(t)} \longleftarrow \theta_{\beta,l}^{(t-1)} + \frac{\tau_t}{2}\nabla \log \pi(\theta_{\beta,l}^{(t-1)})+\frac{\tau_t}{2}\frac{n}{n_s}\nabla\log\pi_{i\in \mathcal{I}}\left(Y_i\mid X_i,Z_i, \theta^{(t-1)}\right)+\sqrt{\tau_t}\varepsilon_l
\end{equation}
where $\varepsilon_l\overset{\mathrm{iid}}{\sim}N(0,1)$. Let $L_r$ be the number of basis coefficients for the $r$-th brain region. The time complexity for \eqref{eq:sgld} is $\min\br{O(L_r^3), O(L_r^2n_s)}$. The full sample size $n$ is usually significantly larger than $L_r$, hence, approximating the full likelihood with a subsample of the data significantly decreases the computational complexity.

Based on the mini-batch idea of the SGLD algorithm, we propose the following Algorithm \ref{algo1}, where the large data set is first split into $B$ smaller batches. Each batch of data is loaded using memory-mapping techniques. Within each batch, a small subsample of size $n_s$ is randomly drawn to be used at each iteration. By splitting the full data into $B$ batches, we reduce the auxiliary space complexity for computing the required summary statistics down to the size of $B$, instead of the size of the full data.

\begin{algorithm}[ht!]
\caption{Scalable Bayesian Image-on-Scalar~(SBIOS) regression with memory mapping}
\label{algo1}
\begin{algorithmic}[1]
\State Set subsample size $s$, an integer $t_I$ for the frequency to update $\eta_i$, and initial values for all parameters. 
\State Split the entire sample into $B$ batches, sequentially load each batch of data and save each batch to the disk using memory-mapping. Set batch index $b$ to 1.
\For{ iteration $t=1,2,\ldots, T$} 
    \State Update the step size $\tau_t$ for $t$-th iteration.
    \State Load batch $b$ into memory.
    \For{ region $r=1,...,R$}
       \State Randomly select a subsample $\mathcal{I}$ of size $n_s$ from batch $b$.
       \State Update $\theta_{\beta,l}$ for region $r$ using \eqref{eq:sgld} based on the selected subsample.
    \EndFor
    \State Update $\gamma,\delta,\sigma_\gamma,\sigma_\beta$ using Gibbs sampling. \label{algo1:line:gamma}
    \State $b=b+1$. If $b>B$, set $b=1$.
    \If {$t$ is a multiple of $t_I$}
        \State Iterate through all batches to update $\{\eta_i\}_{i=1}^N$. \label{algo1:line:eta}
        \State Update $\sigma_Y, \sigma_\eta$ using Gibbs sampling.
        \State (Optional) Impute the missing outcome $Y_i(s_j)$ for all the missing voxel indices $s_j\notin \cV_i$.\label{algo1:line:Y_imp}
    \EndIf
\EndFor
\end{algorithmic}
\end{algorithm}
On line \ref{algo1:line:gamma} of Algorithm \ref{algo1}, using Gibbs sampling to update $\gamma,\delta$ also requires the entire data set, but the posterior distributions of $\gamma$ and $\delta$, in fact, only rely on summary statistics that can be pre-computed based on the entire data set. In practice, at the beginning of the algorithm, we iterate through each batch once to compute these summary statistics and directly use them to update $\gamma$ and $\delta$ at each iteration. The same cannot be done for $\theta_{\beta,l}$, because $\delta(s)$ can be 0 or 1, and the posterior variance for 
 $\left(\theta_{\beta,1},\dots,\theta_{\beta,L}\right)^T$ depends on $Q^T\diag\br{\delta}Q$, which is no longer a diagonal matrix and requires updating at each iteration. See Equation \eqref{eq:working_model}. Hence, sampling $\theta_{\beta,l}$ is more computationally demanding than sampling $\theta_{\gamma,l}$. We provide detailed derivations of the posterior distribution for $\theta_{\beta,l}, \theta_{\gamma,l}, \delta(s)$ in the Supplementary Material, Section 1.

On line \ref{algo1:line:eta} of Algorithm \ref{algo1}, since $\theta_{\eta,i,l}$ is the coefficient for individual-level effect, we must iterate through all samples to update all of the $\theta_{\eta,i,l}$. As such, storing $\theta_{\eta,i,l}$ in memory grows as the sample size $n$ increases. As $\eta_i(s)$ is more of a nuisance parameter and not our main focus, we choose to update $\eta_i$ less frequently. Further improvements to memory allocation can be implemented if we also use batch-splitting on $\eta$ and save the samples of $\eta$ as file-backed matrices. 

For the optional imputation found on line \ref{algo1:line:Y_imp} of Algorithm \ref{algo1}, imputing $Y_i(s_j)$ where $s_j\notin\cV_i$ means that the missing values in $Y_i$ and all summary statistics associated with $Y_i$ need to be updated every $t_I$ iterations. Hence we use an index-based updating scheme. For each $i$, denote the complementary set $\cV_i^c:= \br{s_j}_{j=1}^p - \cV_i$ as the index set of all missing voxels for individual $i$. We keep track of $\cV_i^c$, and create a vector \texttt{Y\_imp} to store the imputed outcome values only on those indices in $\cV_i^c$ for each $i$, and update the corresponding summary statistics every time $Y_i(s_j), s_j\in \cV_i^c$ is updated. The time complexity to update all missing values in $Y_i$ is $O(L\times (|\cV_i^c|))$ where $|\cV_i^c|$ is the number of missing voxels for individual $i$, and $L$ is the total number of basis coefficients. We provide a detailed algorithm for updating missing outcomes in the Supplementary Materials.

Algorithm \ref{algo1} is implemented using the Rcpp package \texttt{bigmemory} \citep{bigmemory}.  The \texttt{bigmemory} package allows us to store large matrices on disk as a \texttt{big.matrix} class and extract the address of the large matrices. At the beginning of Algorithm \ref{algo1}, when the entire data is split into smaller batches, each batch is loaded in R as a \texttt{big.matrix} class, then the address of these big matrices is passed to Algorithm \ref{algo1}, accessing different batches of data becomes very efficient and memory-conserving.

\subsection{Evaluation Criteria}

To assess the variable selection accuracy, we compute the True Positive Rate (TPR) when the False Positive Rate (FPR) is controlled at $10\%$. Because of the selection variable $\delta(s)$, we can obtain a Posterior Inclusion Probability (PIP) from the $m=1,\dots, M$ posterior MCMC samples of $\delta(s)$, 
\[\text{PIP}(s) = \frac{1}{M}\sum_{m=1}^M\delta_m(s).\]
Setting a threshold between 0 and 1 on $\text{PIP}(s)$ gives a mapping of the active pixels. Hence we choose 20 evenly-spaced points between 0 and 1 to fit an ROC curve with linear splines, and get the estimated TPR when FPR is at 10\% from the fitted ROC curve.
For MUA, we choose the cutoff on p-values to be the 20 quantiles of the Benjamini-Hochberg (BH) adjusted p-values corresponding to the probabilities at the 20 evenly-space points, and use the same method to obtain TPR when FPR controlled at 10\%.

All three Bayesian methods (BIOS, SBIOS0, SBIOSimp) are implemented using Rcpp \citep{Rcpp} with RcppArmadillo \citep{RcppArmadillo}\footnote{Code for all simulations and all implementations of the proposed methods and competing methods can be found in the R package SBIOS (\url{https://github.com/yuliangxu/SBIOS}). }.

\subsection{Ablation Study Design}

To demonstrate the performance of the proposed imputation and computation schemes, in Supplementary Section~\ref{sec:simulation}, we compare our proposed model in three variations, BIOS, SBIOS0, and SBIOSimp, with the existing method MUA in three simulation examples. Note that BIOS, SBIOS0, and SBIOSimp are based on the same model \eqref{eq:model} with the same set of priors. The difference is that, BIOS uses a fully Gibbs-sampler without the advanced computational techniques (SGLD, memory-mapping, etc.) with 0 imputation; SBIOS0 uses advanced computation techniques with 0 imputation; and SBIOSimp uses advanced techniques with PCA-based imputation (which we refer to as Gibbs-sampler imputation approach).  From BIOS, SBIOS0, to SBIOSimp, we added one advanced feature at a time as shown in Figure \ref{fig:diff_sbios}. We implemented and compared all three methods with the baseline method MUA in Simulations I to III in Supplementary Section~\ref{sec:simulation}, serving as an ablation study to systematically remove or modify components of an algorithm and understand the contribution of each component. 
\begin{figure}[h]
    \centering    
    \begin{tikzpicture}[->, node distance=6cm, thick]
        \node (A) [rectangle, draw, minimum width=1cm, minimum height=0.8cm] {BIOS};
        \node (B) [rectangle, draw, minimum width=1cm, minimum height=0.8cm, right of=A] {SBIOS0};
        \node (C) [rectangle, draw, minimum width=1cm, minimum height=0.8cm, right of=B] {SBIOSimp};

        \draw (A) -- node[above] {+ Scalable computation} (B);
        \draw (B) -- node[above] {+ GS imputation} (C);
    \end{tikzpicture}
    \caption{Incremental Differences of BIOS, SBIOS0, and SBIOSimp.}
    \label{fig:diff_sbios}
\end{figure}
The three simulation results in Supplementary Section~\ref{sec:simulation} show the superior performance of SBIOSimp in three aspects: selection accuracy (Simulation I), maximum memory usage (Simulation II), and time scalability (Simulation III).

\section{UK Biobank Application}\label{sec:RDA}

In this real data application, we use 3D task fMRI data from the UK Biobank as the outcome and age as the single exposure variable. The three confounding variables are gender, age by gender interaction, and head size. The original range of age is from 44 to 83. We use the standardized age (standardized by $\{X_i-\bar{X}\}/\text{SD}(X)$ where $\bar{X}$ and $\text{SD}(X)$ stands for the sample mean and standard deviation respectively) as the exposure. Gender is coded as a binary variable with 0 being female and 1 being male. The interaction of age by gender is computed using the standardized age times gender. The parameter for the interaction term, $\gamma_{\text{age}\times \text{gender}}(s)$, represents the standardized age effect for males minus the age effect for females.

\subsection{Data Preprocessing and Estimation Procedure}\label{subsec:RDA_preprocess}

The outcome variable is the fMRI data obtained from an emotion recognition task. In this task, participants are required to identify which of two faces displaying fearful expressions (or shapes) presented at the bottom of the screen matches the face (or shape) displayed at the top of the screen (See details in the Hariri faces/shapes emotion task \citep{Miller2016-tq, Hariri2002-hj}, as implemented in the Human Connectome Project). The 3D fMRI data in MNI atlas space \citep{evans1994mri} is a rectangular prism of $91\times 109\times 91$ voxels, and we used a total of $n=38{,}639$ subjects with task fMRI. The NIFTI data is approximately 180 GB, and the processed RDS files are roughly 34 GB. The original NIFTI outcome data for one subject is around 3Mb, and the NIFTI mask data of binary format for one subject is around 26Kb. Using the R package \texttt{RNifti} \citep{RNifti}, the preprocessing time for one subject's data, including directly loading the outcome and mask data from a DropBox folder, is around 2.3 seconds. Saving the preprocessed data into file-backed matrices is instantaneous. We use the Harvard-Oxford cortical and subcortical structural atlases \citep{Desikan2006-wt} for the analysis. After preprocessing, we have a total of 110 brain regions.

\begin{figure}[!ht]
\centering
\includegraphics[width=0.7\textwidth]{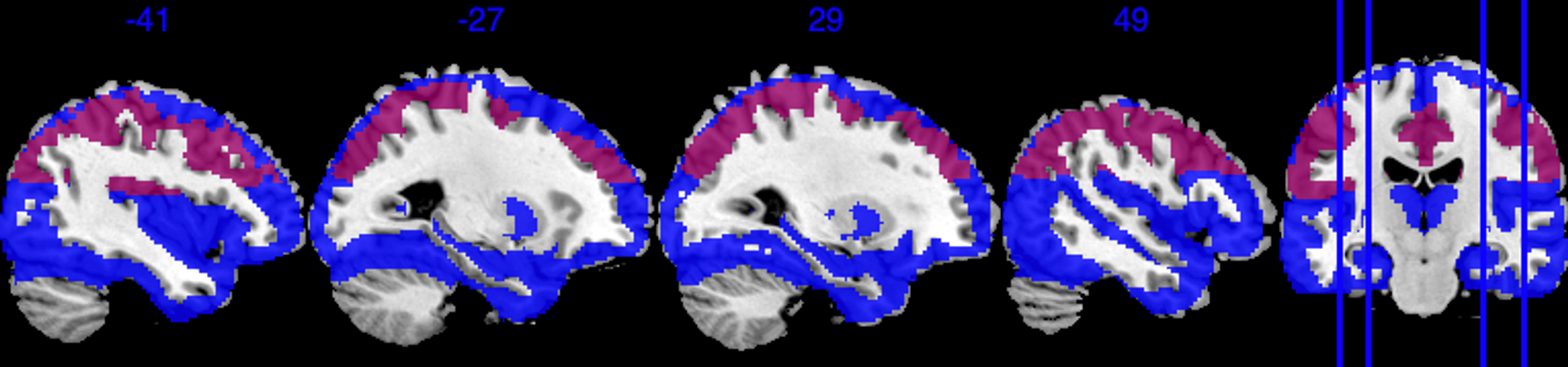}
\caption{Analysis mask using an observed proportion threshold of 0.5 and an intersection mask (completely observed data). The purple area indicates 100\% inclusion; the blue area indicates the mask with an observed proportion between 0.5 and 1.0. }
\label{fig:dropout_50v100}
\end{figure}

 To define the analysis mask, we recall the definition of observed proportion (OP) at location $s_j$ to be $h(s_j) = n^{-1}\sum_{i=1}^n \indi_{\cV_i}(s_j)$, where $\cV_i$ is the set of all observed locations for individual $i$. We define the group analysis mask as $\cB = \br{s_j: h(s_j)>0.5}$, i.e., the area where each voxel has at least 50\% observed data. As shown in Figure \ref{fig:dropout_50v100}, the group analysis mask with completely observed data (purple area) covers significantly less area compared to $\cB$, which has at least 0.5 observed proportion (blue area). In the complete observed data,  large portions of the brain regions are missing, notably including the orbitofrontal cortex, the inferior temporal cortex, and the amygdala—regions crucial for emotion processing. In particular, the mask with complete observed data contains only 52 out of the 110 regions in the Harvard-Oxford atlas.

After applying a common mask $\cB$ with an observed proportion of 0.5, we end up with $\cB$ that contains $p=121{,}865$ voxels. The image outcomes $Y_i(s)$ are standardized across subjects, i.e. $Y_i(s) = \{M_i(s)-\bar M(s)\}/\text{SD[M(s)]}$ where $M_i(s)$ is the original image for subject $i$ location $s$, $\bar M(s)$ is the sample mean, and $\text{SD[M(s)]}$ is the standard deviation of $\{M_i(s)\}_{i=1}^n$\phantomsection\label{sent:standardization}. For each region, we apply the Mat\'ern kernel function but with different $\rho$ and $\nu$ parameters \eqref{eq:matern}, to account for the different smoothness of each region. Both $\rho$ and $\nu$ are determined through grid search so that the empirical covariance of $Y(s_j),Y(s_j')$ and the estimated covariance by the Mat\'ern kernel have the smallest difference in Frobenius norm.  The number of bases $L$ is chosen so that the cumulative summation $\sum_{l=1}^L\lambda_l$ accounts for 90\% of the total summation of all eigenvalues, hence we have a total number of $L=\sum_{r=1}^{110} L_r=16{,}879$. In Section \ref{sec:RDA_sensi}, we provide a sensitivity analysis when the cutoff is based on 92\% of the total summation.

The $n=38{,}639$ subjects are split into 50 batches of data, with each batch containing around $700$ to $800$ subjects. The subsample size is 200, and the step size decay parameters $a=0.0001,b=1,\gamma=0.35$ are chosen so that the step size roughly decreases from $7\times 10^{-5}$ to $5\times 10^{-6}$ over the 5,000 MCMC iterations. We use the MUA results as the initial values for $\beta$ and $\gamma$ to run the SBIOS0 and SBIOSimp methods.
 The initial values for $\theta_{\eta,i,l}$ are set to 0 everywhere, and the initial values for $\delta(s)$ are set to 1 everywhere. The initial values for $\sigma_Y,\sigma_\eta,\sigma_\beta,\sigma_\gamma$ are all set to  0.1.
To check the convergence of SBIOSimp, we run three independent chains and use the Gelman and Rubin test. The $l_2$ norm of the residuals is computed for the last 1,000 iterations from the three chains using 20\% of the entire data. The point estimate of the Gelman-Rubin statistic is 1.03 with an upper confidence limit of 1.1,
indicating that the MCMC chain has approximately converged.

\subsection{Analysis Results}
\subsubsection{Age-Related Emotion Recognition Brain Activation Patterns }\label{sec:RDA_result}
We present the top 10 regions identified by the SBIOSimp method in Table \ref{tb:RDA_region}. This SBIOSimp result takes 20.5 hours to run on 1 core CPU HPC Cluster, and the max memory used is 14 GiB.
Using the posterior sample for each region, we compute the Region Level Activation Rate (RLAR) as follows: denote $\mathcal{B}_j$ as the mask for region $j$, the RLAR is $\sum_{s\in \mathcal{B}_j}\delta(s)/|\mathcal{B}_j|$, where $|\mathcal{B}_j|$ is the total number of voxels in region $j$. Hence for each MCMC sample of $\delta(s)$, we obtain one sample of RLAR for all regions, and therefore obtain the posterior distribution of RLAR for all regions. We present histograms of the posterior distribution of the RLAR over the last 1,000 MCMC iterations in Section~\ref{sec:supp_RDA} in the Supplementary Material. Table \ref{tb:RDA_region} presents the top 10 regions with the highest posterior mean of RLAR and their 95\% credible intervals. To present the marginal effect of each $\beta(s_j)$, we also report the effect summing over each region computed separately for the positive and negative effects. Only voxels with a marginal posterior inclusion probability (PIP) over 95\% are viewed as active voxels. Because the top 10 regions have no active voxels with positive effects, we only report the negative effect in Table \ref{tb:RDA_region}. We also report the number of active voxels in Table \ref{tb:RDA_region}. All active voxels in the top 10 regions have a negative effect, that is, as age increases, the brain signal intensity on selected voxels tends to decrease. This trend is also reflected in the raw data, as can be seen by the scatter plots of age against the average image intensity in Section~\ref{sec:supp_RDA} in the Supplementary Material. To interpret the numeric results in Table \ref{tb:RDA_region}, take the \textit{Right temporal fusiform cortex, anterior division} region as an example. On average, 99\% of the voxels within this region have activity that is associated with age.  Specifically, there is a decrease in brain signal intensity of 14.81 standard deviations summed over all locations within this region for 1 standard deviation increase in age. The last column represents the median percentage decline in the brain signal intensity if age increases from 50 to 60. For \textit{Right temporal fusiform cortex, anterior division}, this means when age increases from 50 to 60, the median decline in the brain signal intensity among all voxels in this region will be 75.87\%. Supplementary Section~\ref{supp_sec:percentage} provides the mathematical definition and detailed derivation for this percentage.

\begin{table}[h!]

\centering
\caption{Top 10 regions ordered by Region Level Activation Rate (RLAR). The 4th columns reports the negative voxel effect size summed over each region i.e. ~$\sum_{j\in\cB_j}\mathbb{E}\br{\beta(s_j)|\beta(s_j)<0}$, and inside the bracket are the number of negative voxels. The last column is the percentage changes in the brain intensity when age increases from 50 to 60 (see Section~\ref{supp_sec:percentage}). Only voxels with marginal inclusion probability greater than 0.95 are included, otherwise counted as zero effect voxel. Hence positive voxels are omitted due to low inclusion probability.}
\label{tb:RDA_region}
\resizebox{\columnwidth}{!}{
\begin{tabular}{lccccc}
    \hline
    Region Name & Size & RLAR & Neg Sum (Count) &  Median 50-60  \\
    \hline
Right intracalcarine cortex                       & 634  & 1    & -47.08 (632)    & -53.83          \\
Right supracalcarine cortex                       & 151  & 1    & -8.38 (151)     & -52.15           \\
Left temporal fusiform cortex, anterior division  & 301  & 1    & -15.57 (299)    & -81.18           \\
Left inferior frontal gyrus, pars triangularis    & 692  & 1    & -48.86 (686)    & -56.38            \\
Right inferior temporal gyrus, anterior division  & 314  & 1    & -16.54 (303)    & -93.38           \\
Left intracalcarine cortex                        & 557  & 0.99 & -40.29 (547)    & -51.38          \\
Right temporal fusiform cortex, anterior division & 276  & 0.99 & -14.81 (263)    & -75.87           \\
Left occipital pole                               & 1977 & 0.99 & -157.61 (1911)  & -57.67        \\
left hippocampus                                  & 218  & 0.98 & -12.51 (209)    & -77.37          \\
Left inferior temporal gyrus, anterior division   & 346  & 0.98 & -15.26 (316)    & -106.83           \\
    \hline
  \end{tabular}
}
\end{table}

\begin{figure}[ht!]
  \centering
  \begin{tabular}{c}
    \begin{subfigure}{0.8\textwidth}
      \centering
      \includegraphics[width=\textwidth]{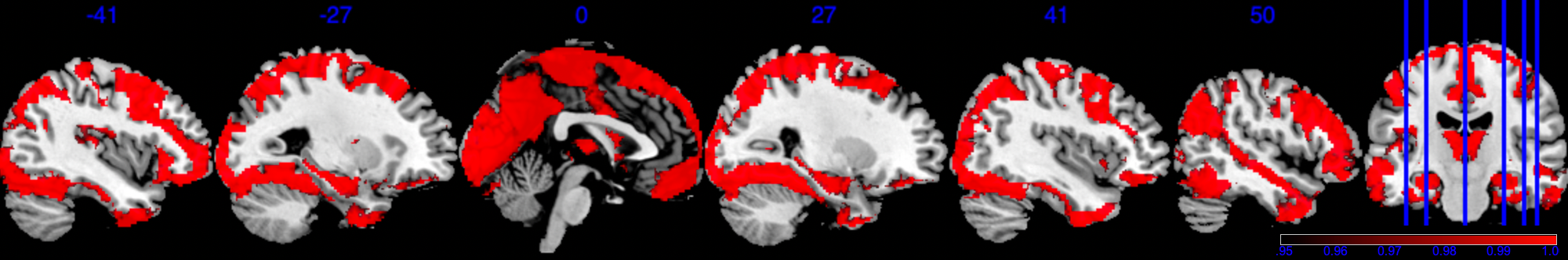}
        \caption{Posterior inclusion probability~(PIP). The color bar from black to red ranges in $[0.95,1]$. Sagittal plane. The first two sagittal slices are in the left hemisphere.}
        \label{fig:RDA_pip}
 \end{subfigure} \\
    
    \begin{subfigure}{0.8\textwidth}
      \centering
      \includegraphics[width=\textwidth]{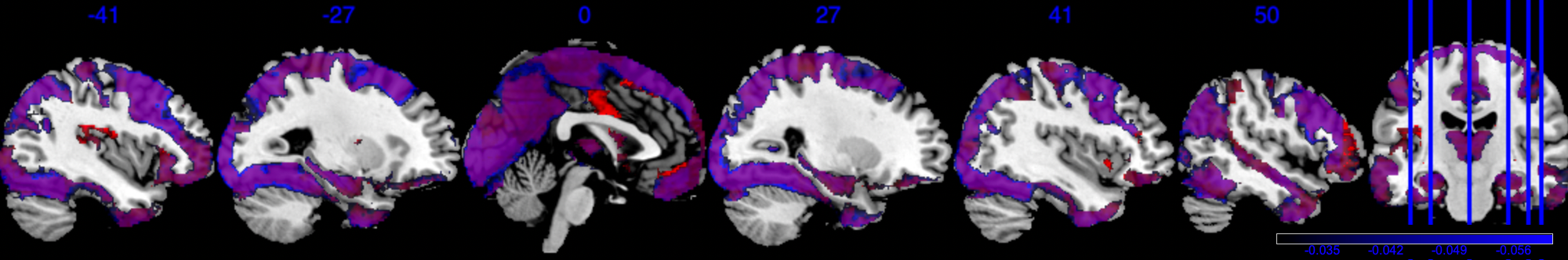} 
        \caption{Posterior mean of  $\beta(s)\delta(s)$. The color bar from black to blue ranges from -0.03 to -0.06. The overlaying purple area is the mask of PIP greater than 0.95. }
        \label{fig:RDA_mean}
    \end{subfigure}
  \end{tabular}
  \caption{Illustration of age-related activation patterns  using a grayscale brain background image (\texttt{ch2bet}, \cite{holmes1998enhancement}). Images are created using MRIcron \citep{rorden2000stereotaxic}.}
  \label{fig:UKB_RDA}
\end{figure}

For a visual representation, Figure \ref{fig:RDA_pip} shows the voxel level PIP in the sagittal plane. The highlighted red regions represent voxels with greater than 0.95 PIP. Figure \ref{fig:RDA_mean} presents the effect size of $\beta(s)\delta(s)$, with the highlighted area in the range $(-0.06, -0.03)$. Note that from Figure \ref{fig:RDA_mean},  voxels with PIP greater than 0.95 also correspond to voxels with a larger absolute value of effect size.
We notice that the activation region (defined by voxel level PIP greater than 0.95) has a negative effect $\beta(s)$. This can also be validated by the scatter plot in Section~\ref{sec:supp_RDA} in the Supplementary Material, where the image intensity generally has a negative association with age across all individuals.

Based on our results, we have the following general interpretations: (i) when controlling for the confounders, age has a negative impact on the neural activity for emotion-related tasks; (ii) the negative effect reflected from each voxel is of very small scale, shown as in Figure \ref{fig:RDA_mean}, indicating a very low voxel level signal-to-noise ratio; (iii) the top 5 brain regions with the highest RLAR are (a) \textit{right intracalcarine cortex}, \textit{right supracalcarine cortex}, and \textit{left Temporal fusiform cortex, anterior division}, all considered as  critical areas for high-level visual processing  including face recognition; 
(b) \textit{left temporal fusiform cortex, anterior division}, a key structure for face perception, object recognition, and language processing \citep{weiner2016anatomical};
and (c) \textit{right inferior temporal gyrus, anterior division}, an area for language and semantic memory processing, visual perception, and multimodal sensory integration \citep{onitsuka2004middle}.
These top 5 regions are also consistently identified in the sensitivity analysis when using half of the data as training data, see Section \ref{sec:RDA_sensi}.

To further demonstrate the voxel-level results provided by SBIOSimp, in the next section, we perform a detailed  analysis of  the amygdala region.

\subsubsection{Focused Results: Amygdala Region }\label{sec:supp_amygdala}

\begin{figure}[htbp]
  \centering
  \begin{subfigure}[b]{0.9\textwidth}
    \includegraphics[width=\textwidth]{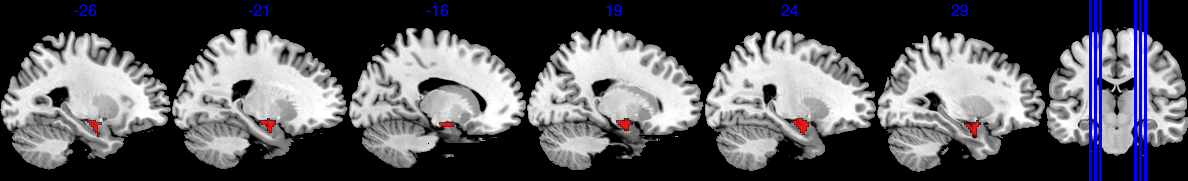}
    \caption{Amygdala (red).}
    \label{fig:amygdala}
  \end{subfigure}

  \begin{subfigure}[b]{0.9\textwidth}
    \includegraphics[width=\textwidth]{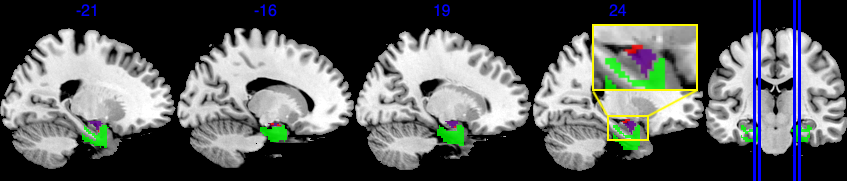}
    \caption{Purple area indicates voxels in the amygdala with $\text{PIP}>0.95$. Green area is the \textit{parahippocampal gyrus, anterior division}}
    \label{fig:amygdala_PGA_IP}
  \end{subfigure}

   \begin{subfigure}[b]{0.9\textwidth}
    \includegraphics[width=\textwidth]{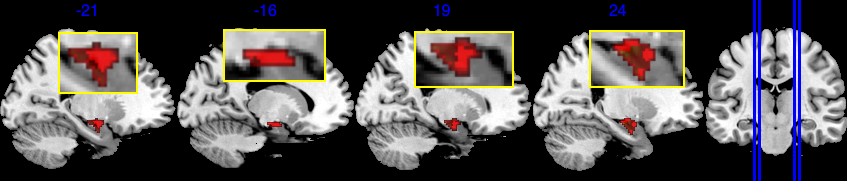}
    \caption{Yellow shaded area indicates voxels in the amygdala with at least 50\% decline in the brain signal intensity for 10-year increase in age from 50.}
    \label{fig:amygdala_age5060_neg100to50}
  \end{subfigure}

  \caption{Illustrations on the amygdala region}
  \label{fig:amygdala_all}
\end{figure}

The amygdala region is part of the limbic system. They are responsible for detecting danger and negative emotions, and play an important role in behavior, emotional control, and learning \citep{bzdok2013investigation}. The emotion task fMRI data in UK Biobank is based on emotion tasks where participants are asked to identify faces with negative emotions. Hence we expect the amygdala region to play an important role. The amygdala region is a small area in the brain, as shown in Figure \ref{fig:amygdala}, containing 380 voxels out of a total of 121,865 voxels.

Based on the results shown in Figure \ref{fig:amygdala_all}, SBIOSimp identifies a large proportion of voxels in the amygdala region to be active. Numerically, SBIOSimp identifies 324 out of 380 voxels in both the left and right amygdala to be active with $\text{PIP}>0.95$. From the highlighted box in Figure \ref{fig:amygdala_PGA_IP}, the active voxels mostly concentrate on the direction where the amygdala region connects to the \textit{parahippocampal gyrus, anterior division}. The anterior portion of the parahippocampal gyrus is involved in complex emotive processes and has significant interconnectivity with other cortical limbic structures and the amygdala \citep{kaas2016evolution}. Figure \ref{fig:amygdala_age5060_neg100to50} shows the yellow-shaded area within the amygdala in which voxels are associated with at least 50\% decline in the brain signal intensity for 10-year increase in age from 50.

\subsection{Comparing SBIOSimp with SBIOS0 and MUA}
\begin{figure}[!ht]
    \centering
    \includegraphics[width=1\textwidth]{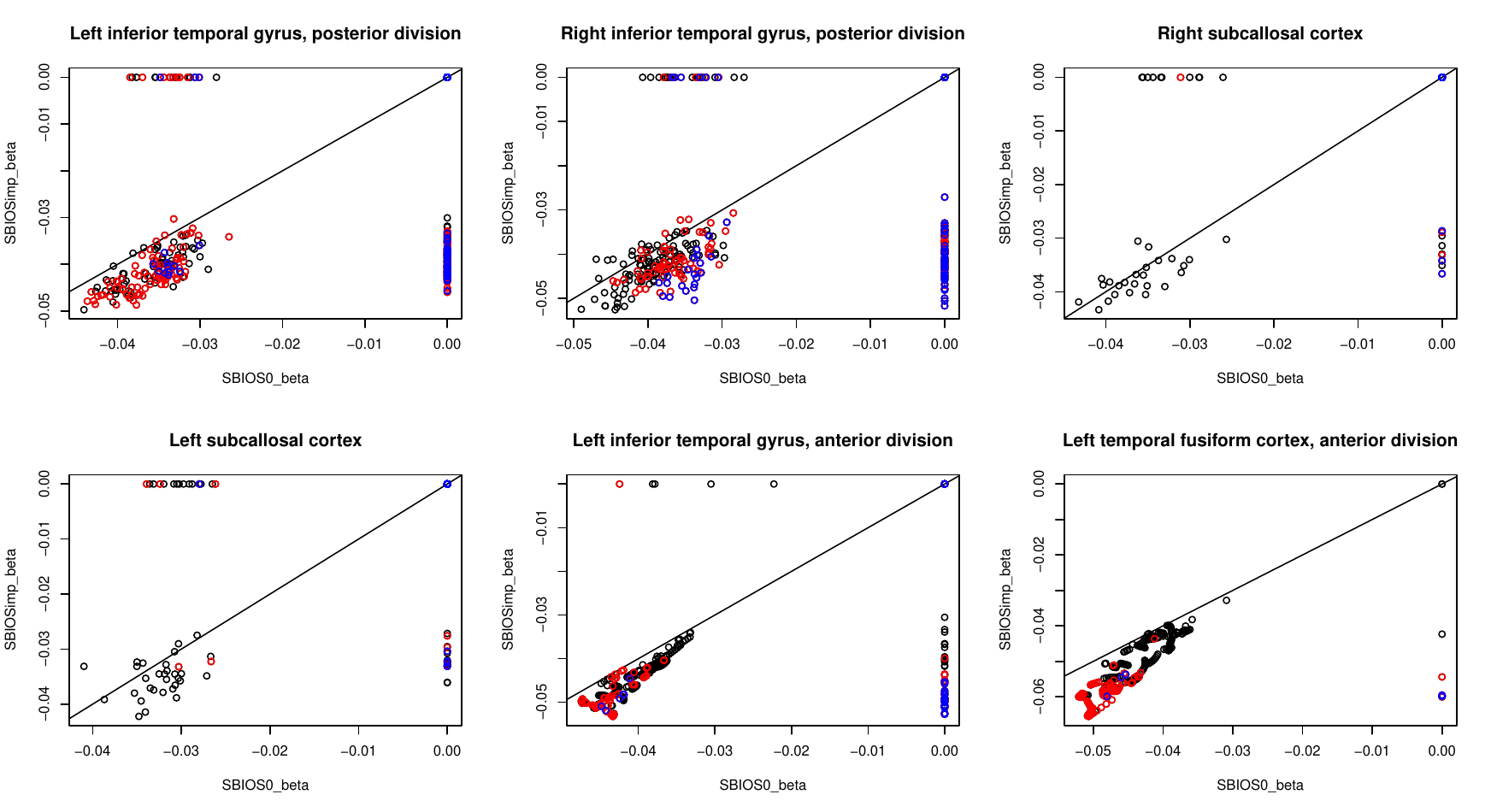}
  \caption{Scatter plot of the posterior mean of $\beta(s_j)I(\text{PIP($s_j$)}\geq 0.95)$ based on SBIOS0 (x-axis) and SBIOSimp (y-axis) on six selected regions with high missingness. Blue dots indicate voxels with observed proportion $h(s_j)\in [0.5,0.7)$. Red dots indicate voxels with observed proportion $h(s_j)\in [0.7,0.9)$.  Black dots indicate voxels with observed proportion $h(s_j)\in [0.9,1]$.}
  \label{fig:compare_beta_imp}
\end{figure}

We compare the posterior mean of $\beta(s)$ given $\text{PIP}(s)\geq 0.95$ between SBIOS0 and SBIOSimp stratified by the observed proportions on these regions in Figure \ref{fig:compare_beta_imp}. Each point in Figure \ref{fig:compare_beta_imp} represents the effect of age on the brain signal for one voxel in the brain. The 6 regions in Figure \ref{fig:compare_beta_imp} are chosen with high missingness. Comparing blue dots (low observed proportion, $h(s_j)\in [0.5,0.7)$), red dots (medium observed proportion, $h(s_j)\in [0.7,0.9)$) with black dots (high observed proportion, $h(s_j)\in [0.9,1]$), we can see that $\beta$ fitted with SBIOS0 on voxels with lower observed proportion tend to be closer to 0 or directly mapped to 0 according to $I(\text{PIP}(s)\geq 0.95)$, compared with SBIOSimp. This implies that by directly imputing missing outcomes with 0, SBIOS0 tends to put more shrinkage on the posterior mean of $\beta(s)$ compared to SBIOSimp. Hence SBIOS0 potentially has lower power than SBIOSimp to detect the signals, which could be justified by the simulation result shown in Figure~\ref{fig:sim_TPR}. 

A comparison of active $\beta$ selection between MUA and SBIOSimp is available in Table \ref{tb:compare_mua}. For SBIOSimp, the active voxel selection is based on PIP greater than 95\%. For MUA, for a fair comparison, the cutoff on BH-adjusted p-values is determined using the same proportion of active voxels selected by SBIOSimp, so that MUA and SBIOSimp select the same number of active voxels. Table \ref{tb:compare_mua} shows that MUA with 0-imputation tends to map more voxels of low observed proportion towards 0.

\begin{table}[ht!]
        \centering
        \caption{Proportion of active voxels  identified by MUA and SBIOSimp for each range of observed proportions (OP).}
        \label{tb:compare_mua}
        \resizebox{0.5\columnwidth}{!}{
        \begin{tabular}{llll}        
        \hline\hline
        OP& {[}0.5, 0.7) & {[}0.7, 0.9) & {[}0.9, 1{]} \\\midrule
        MUA      & 0.15         & 0.39         & 0.62         \\
        SBIOSimp & 0.27         & 0.46         & 0.62        \\
        \hline\hline
        \end{tabular}
        }
        
    \end{table}

\subsection{Sensitivity Analysis}
\subsubsection{Hyperparameter Specifications}\label{sec:RDA_sensi}

To further validate the result reported in subsection \ref{sec:RDA_result}, we conduct sensitivity analysis on different choices of the hyperparameter $IG(a,b)$ in the prior for $\sigma_Y^2$, the choice of $\sigma_\beta^2$, and the number of bases in the Gaussian kernel. 
The baseline setting for results in subsection \ref{sec:RDA_result} is $\sigma^2_Y\sim IG(0.1,0.1)$, hyperparameter $\sigma_{\beta}^2 = 0.01$, and the number of basis is based on 90\% of total eigenvalues ($L=16{,}879$). In the sensitivity analysis, in case 1, the prior for $\sigma_Y^2$ is $IG(1,1)$, and $IG(0.1,0.1)$ for case 2. In case 3, we use $\sigma_{\beta}^2 = 1$. In case 4, we choose the number of bases based on 92\% of total eigenvalues ($L = 20{,}355$). For case 5, we vary the smoothness parameter $\nu$ in the Mat\'ern kernel \eqref{eq:matern}, and create a new kernel where $\nu$ in each region is set to 90\% the value of $\nu$ in the standard kernel used in Section \ref{sec:RDA_result}.
Because Case 1 \& 2 have very similar results, we report them together.

Table \ref{tb:sensi} presents the region-level difference between each sensitivity case and the final result presented in subsection \ref{sec:RDA_result}. Cases 1 and 2 produce almost the same result and are reported together. We can see that the top four regions are consistently selected with the highest RLAR.

\begin{table}[h]
\centering
\caption{Sensitivity Analysis of SBIOSimp on UK Biobank data. The RLAR for each region is reported. For the region names, (R) means right hemisphere, (L) means left hemisphere, AD means anterior division,  PT means pars triangularis.}
\label{tb:sensi}
\resizebox{0.9\columnwidth}{!}{
\begin{tabular}{@{}cccccc@{}}
\hline
Region ID & Region Name                                       & Cases 1\&2 & Case 3  & Case 4 & Case 5  \\
\hline
24        & (R) intracalcarine cortex                       & 1.00          & 1.00      & 1.00      & 1.00      \\
47        & (R) supracalcarine cortex                       & 1.00          & 1.00      & 1.00      & 1.00      \\
85        & (L) temporal fusiform cortex, AD                       & 1.00          & 1.00      & 1.00      & 1.00      \\
53        & (L) inferior frontal gyrus, PT                         & 1.00          & 1.00      & 1.00      & 1.00      \\
14        & (R) inferior temporal gyrus, AD  & 1.00          & 1.00      & 0.99   & 0.99   \\
72        & (L) intracalcarine cortex                        & 0.99       & 0.99   & 0.99   & 0.99   \\
37        & (R) temporal fusiform cortex, AD & 0.99       & 0.99   & 0.99   & 1.00      \\
96        & (L) occipital pole                               & 0.99       & 0.99   & 0.99   & 0.99   \\
105       & (L) hippocampus                                  & 0.98       & 0.99   & 0.98   & 0.99   \\
62        & (L) inferior temporal gyrus, AD   & 0.98       & 0.98   & 0.99   & 0.99   \\
\hline
\end{tabular}%
}
\end{table}

\subsubsection{Model assumptions on selection of interactions}\label{supp_sec:RDA_delta}

In the analysis of Section \ref{sec:RDA_result}, we apply the selection variable $\delta(s)$ exclusively to the main effect of age. In this section, we extend the model \eqref{eq:model} by applying $\delta(s)$ to both the main effect of age and the interaction effect of age and gender. Consequently, the interaction term is automatically set to zero whenever the corresponding main effect is not selected. The extended model can be expressed as below:
\begin{equation}
    Y_i(s_j) = X_i \beta(s_j)\delta(s_j) + \tilde X_i \tilde \beta(s_j)\delta(s_j) + \sum_{k=1}^m\gamma_k(s_j)Z_{ik} + \eta_i(s_j) + \epsilon_i(s_j), \label{eq:model_delta}
\end{equation}
where $\epsilon_i(s_j) \sim \mathrm{N}(0,\sigma_Y^2)$ and $X_i$ is the standardized age for individual $i$, and $\tilde X_i$ is the interaction term between age and gender. In this way, $\delta(s)$ can control the active signals in the main effect and interaction effect at the same time. Here, $\beta$ and $\tilde\beta$ are assigned GP priors independently and both are updated using the SGLD algorithm. 

\begin{table}[h!]

\centering
\caption{Region level results when applying $\delta$ to both age and the interaction of age and gender, on the same regions as reported in Table \ref{tb:RDA_region}. See the caption of  Table \ref{tb:RDA_region} for the interpretation of column names.}
\label{tb:RDA_region_delta}
\resizebox{0.9\columnwidth}{!}{
\begin{tabular}{ccccc}
\hline
Region Name                      & Size & RLAR             & Neg Sum (Count)  & Neg/Count\\ 
\hline
Right intracalcarine cortex                       & 634  & 0.95 & -39.38 (521)   & -0.076 \\
Right supracalcarine cortex                       & 151  & 1.00    & -7.04 (151)    & -0.047 \\
Left temporal fusiform cortex, anterior division  & 301  & 1.00    & -15.05 (301)   & -0.050 \\
Left inferior frontal gyrus, pars triangularis    & 692  & 0.99 & -42.64 (654)   & -0.065 \\
Right inferior temporal gyrus, anterior division  & 314  & 1.00    & -14.77 (312)   & -0.047 \\
Left intracalcarine cortex                        & 557  & 0.99 & -35.74 (534)   & -0.067 \\
Right temporal fusiform cortex, anterior division & 276  & 1.00    & -13.84 (275)   & -0.050 \\
Left occipital pole                               & 1977 & 0.84 & -120.70 (1303) & -0.093 \\
left hippocampus                                  & 218  & 0.98 & -11.99 (211)   & -0.057 \\
Left inferior temporal gyrus, anterior division   & 346  & 1.00    & -14.32 (346)   & -0.041 \\
\hline
\end{tabular}
}
\end{table}

The last column in Table \ref{tb:RDA_region_delta} represents the average negative effect among the active voxels, which is just the \{Neg Sum\}/\{Count\} in the previous column. In Table \ref{tb:RDA_region_delta}, there are four regions, \textit{Right intracalcarine cortex}, \textit{Left inferior frontal gyrus, pars triangularis}, \textit{Left intracalcarine cortex}, and \textit{Left occipital pole} have made less active voxel selection after applying a common $\delta$ to the interaction term, whereas the remaining six regions have little changes.

\begin{figure}[ht!]
  \centering
  \begin{tabular}{c}
    \begin{subfigure}{0.8\textwidth}
      \centering
      \includegraphics[width=\textwidth]{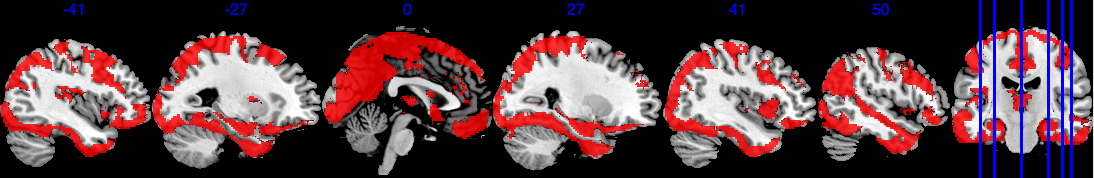}
        \caption{Posterior inclusion probability~(PIP). The color bar from black to red ranges in $[0.95,1]$. Sagittal plane. The first two sagittal slices are in the left hemisphere.}
        \label{fig:RDA_delta_pip}
 \end{subfigure} \\
    
    \begin{subfigure}{0.8\textwidth}
      \centering
      \includegraphics[width=\textwidth]{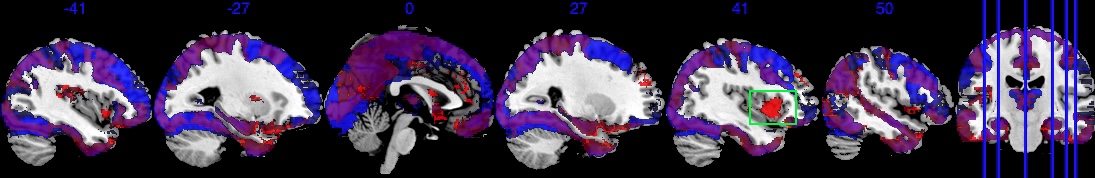} 
        \caption{Posterior mean of  $\beta(s)$. The color bar from black to blue ranges from -0.03 to -0.06. The overlaying purple or red area is the mask of PIP greater than 0.95. }
        \label{fig:RDA_delta_mean}
    \end{subfigure}\\
    \begin{subfigure}{0.8\textwidth}
      \centering
      \includegraphics[width=\textwidth]{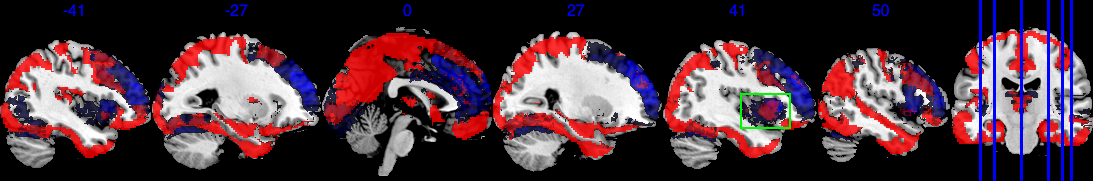} 
        \caption{Posterior mean of  $\tilde\beta(s)$ (positive). The color bar from black to blue ranges from 0.01 to 0.06. The overlaying purple or red area is the mask of PIP greater than 0.95. }
        \label{fig:beta_tilde}
    \end{subfigure}\\
    \begin{subfigure}{0.8\textwidth}
      \centering
      \includegraphics[width=\textwidth]{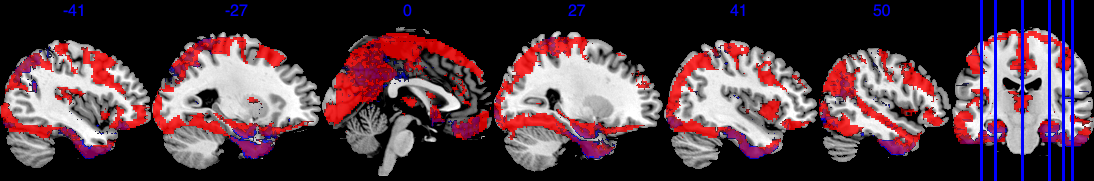} 
        \caption{Posterior mean of  $\tilde\beta(s)$ (negative). The color bar from black to blue ranges from -0.01 to -0.02. The overlaying purple or red area is the mask of PIP greater than 0.95. }
        \label{fig:beta_tilde_neg}
    \end{subfigure}
  \end{tabular}
  \caption{Illustration of results after applying a common $\delta(s)$ term on both the main effect of age and the interaction effect between age and gender, using a grayscale brain background image (\texttt{ch2bet}, \cite{holmes1998enhancement}). Images are created using MRIcron \citep{rorden2000stereotaxic}. The highlighted green boxed area indicates one active region that has small effect size for $\beta(s)$ but large effect size for $\tilde\beta(s)$.}
  \label{fig:UKB_RDA_delta}
\end{figure}

In addition, Figure \ref{fig:UKB_RDA_delta} shows results equivalent to those in Figure ~\ref{fig:UKB_RDA}, but with the selection indicator $\delta(s)$ applied simultaneously to both the main effect and the interaction effect. As shown in Figure~\ref{fig:RDA_delta_mean},  the posterior of $\delta(s)$ in model~\eqref{eq:model_delta} is driven by both $\beta(s)$ and $\tilde \beta(s)$. The regions exhibiting strong negative effects of $\beta(s)$ do not fully align with the regions showing $\text{PIP}>0.95$, in contrast to the original results in Figure \ref{fig:RDA_mean}. 
Therefore, if the primary interest is in the main exposure variable $X_i$, such as age in our analysis, we recommend applying $\delta(s)$ exclusively to this primary effect to achieve a more accurate selection of activation regions.

Since the gender variable is binary with female being 0 and male being 1, the interpretation for $\tilde\beta(s)$ is that comparing to the female subjects, one standard deviation (s.d.) increase in age for male subjects is associated with $\tilde\beta(s)$-s.d. of change in the image intensity. The boxed green area in Figure \ref{fig:beta_tilde} and Figure \ref{fig:RDA_delta_mean} identifies one active area where $\beta(s)$ has a negligible effect, but $\tilde\beta(s)$ has a large effect size, indicating that this area is associated with the differences of male's age-brain intensity association compared to female. For example, one s.d. increase in male's age is associated with at least 0.01 s.d. increase in brain signal intensity compared to the female baseline in this green-boxed area. On the other hand, Figure \ref{fig:beta_tilde_neg} also identifies areas where one s.d. increase in male's age is associated with at least 0.01 s.d. decrease in brain signal intensity compared to the female baseline. The area in the green box spans several brain regions in the right hemisphere, including \textit{Right lateral occipital cortex, superior division}, \textit {Right insular cortex}, \textit{Right middle temporal gyrus, posterior division}, and \textit{Right frontal operculum cortex}. They jointly integrate information from multiple modalities and detect behaviorally relevant stimuli. The negative $\tilde\beta(s)$ in Figure \ref{fig:beta_tilde_neg} spans over \textit{Right parahippocampal gyrus, posterior division, Right temporal fusiform cortex, posterior division, Left temporal fusiform cortex, posterior division, Left temporal pole}. They jointly process and integrate vision and semantic information and are related to contextual and memory functions.

\section{Discussion}\label{sec:discussion}

We propose a Bayesian hierarchical model for Image-on-Scalar regression, with a computationally efficient algorithm for posterior sampling, and apply our proposed method to the UK Biobank data. Our proposed model can capture high dimensional spatial correlation, account for the individual-level correlated noise, and provide uncertainty quantification on the active area selection through posterior inclusion probability. Our main computational contribution includes (i) employing a scalable Stochastic Gradient Langevin Dynamics (SGLD) algorithm for big data  ISR, (ii) borrowing memory-mapping techniques to overcome memory limitations for large-scale imaging data, and (iii) utilizing individual-specific masks with an imputation approach to maximize the use of available imaging data.

Extensive simulations compare our model's implementations — standard Gibbs sampling (BIOS), SGLD with missing outcomes imputed as 0 (SBIOS0), and our imputation-based SGLD (SBIOSimp) — against Mass Univariate Analysis (MUA). Results show that SBIOSimp achieves superior variable selection accuracy, lower memory cost, and scalable computation. UK Biobank application results are validated by sensitivity analysis under different hyperparameters and kernel settings. Note that although we only compare SBIOSimp with the baseline 0-imputation method since 0-imputation is the only off-the-shelf alternative imputation method to be applied on such a large scale, there exist other directions of imputation approaches, such as the functional PCA \citep{happ2018multivariate}\label{sent:fPCA}.

Simulations and UK Biobank data demonstrate our method's potential for large-scale imaging, yet several limitations persist. First, our Gaussian process depends on a user-defined kernel function, and our adaptive kernel parameter selection is partially subjective. Second, for computational efficiency we assume the individual effect $\eta_i$ shares the same kernel as $\beta$ and $\gamma$, which may overlook individual-level latent confounders. Lastly, while SGLD accurately estimates the first moment of $\beta$, its diminishing step size may fail to capture the true variance of $\beta$.

\section*{Acknowledgments}
\vspace{-1em}

\noindent
The authors would like to thank the Editor Professor Michael Stein, the
Associate Editor and reviewers for their helpful comments and constructive
suggestions, which led to this much-improved manuscript. 

\section*{Disclosure Statement}
The authors have no conflicts of interests to declare.

\section*{Funding}
This work was supported by National Institute of Health grant  R01DA048993. Kang's research was partially supported by National Institute of Health grant R01MH105561 and National Science Foundation grant IIS2123777.

\section*{Data Availability Statement}
\vspace{-1em}

\noindent
The real data, UK Biobank data \citep{Miller2016-tq}, used in Section~\ref{sec:RDA}, is publicly available online at \url{https://www.ukbiobank.ac.uk/}. The preprocessed data supporting the findings of this study are available from the corresponding author on request.

\spacingset{1.3} 

\bigskip
\bibliographystyle{asa}
\bibliography{ref}

\appendix

\newpage 

The supplementary material contains three sections. Section \ref{sec:supp_deriv} provides the posterior derivation and the detailed imputation algorithm. Section \ref{sec:simulation} provides the main simulation results. Section \ref{sec:supp_sim} provides additional simulation results. Section \ref{sec:supp_RDA} provides additional real data analysis results. Section \ref{sec:supp_RDA_sim} provides UKBiobank scale simulation studies, whereas Section \ref{sec:supp_oneKer} provides a low-resolution simulation using a single GP kernel.

\section{Full Conditional Posterior Distributions}\label{sec:supp_deriv}
We provide the derivations of full conditional posterior distribution for each parameter. We will consider the case with completely observed data, and then provide the derivation for the case with imputing the missing values when considering individual masks.

We use $I_d$ to denote the identity matrix in $\R^{d \times d}$. We use $\diag\br{D}$ to denote a diagonal matrix whose diagonal vector is $D$.

\subsection{Posterior densities with completely observed data}
Consider the basis expansion approximations of the GPs
\begin{align}
    \beta(s) &\approx \sum_{l=1}^L \theta_{\beta,l}\psi_l(s)\\
    \gamma_k(s) &\approx \sum_{l=1}^L\theta_{\gamma,k,l}\psi_l(s)\\
    \eta_i(s) &\approx \sum_{l=1}^L\theta_{\eta,il}\psi_l(s)
\end{align}
where $\theta_{\beta,l}\overset{ind}{\sim}N(0,\sigma_{\beta}^2\lambda_l)$. Denote $D = \sbr{\lambda_1,\dots,\lambda_L}^T$ as the vector of eigenvalues.

Let $Q\in \R^{p\times L}$ be the basis decomposition matrix, where $Q_{j,l} = \psi_{l}(s_j)$. Here, to approximate the orthonormality of $\psi_l(s)$, $Q^TQ = I_{L}$, $Q$ is made an orthonormal matrix using QR decomposition.

Denote $Y^*_i=Q^T Y_i \in \R^L$ as the low-dimensional mapping of the $i$th image $Y_i\in \R^p$. After basis decomposition
\begin{align}
    Y_i^* &= Q^T X_i\diag\{\delta\} Q\theta_\beta + \sum_{k=1}^{m}\theta_{\gamma,j} Z_{i,k}+\theta_{\eta,i} + \epsilon^*_i\\
    \theta_\beta(s) &\sim  N(0,\sigma_\beta^2 \diag\br{D})\\
    \theta_{\gamma,k}(s) &\sim  N(0,\sigma_\gamma^2 \diag\br{D})\\
    \epsilon^*_i(s) &\sim N(0,I_L)
\end{align}
The posterior distribution can be derived as follows.
\begin{align*}
    \Sigma_{\beta(s)}|\text{rest} & =   \left(\frac{1}{\sigma_\beta^2}\diag\br{D}^{-1} +\sum_{i=1}^n Q^T\diag\left\{X_i\delta\right\}Q
     \left(\frac{1}{\sigma_Y^2}I\right)Q^T\diag\left\{X_i\delta(s)\right\}Q\right)^{-1}\\
     \theta_\beta(s)|\text{rest}  &\sim N\left( \Sigma_{\theta_\beta(s)|\rest}\left\{
    \frac{1}{\sigma_Y^2}\sum_{i=1}^n\left\{
    [Y_i^*-\theta_{\eta,i}- \sum_{k=1}^{m}\theta_{\gamma,k} Z_{i,k}]^TQ^T\diag\{X_i\delta(s)\}Q
    \right\}
    \right\} ,\Sigma_{\theta_\beta(s)|\rest}\right)\\
\Sigma_{\gamma_k(s)}|\text{rest} &=  \left(\frac{1}{\sigma_\gamma^2}D^{-1} +\sum_{i=1}^n \frac{1}{\sigma_Y^2}Z_{i,k}^2I_L\right)^{-1}\\
\theta_{\gamma,k}(s)|\text{rest}  &\sim N\left( \Sigma_{\theta_{\gamma_j}(s)|\rest}\left\{
    \frac{1}{\sigma_Y^2}\sum_{i=1}^n Z_{k,i}\left\{
    Y_i^*-\theta_{\eta,i}- \sum_{k'\neq k}\theta_{\gamma,j} Z_{k',i} - Q^T X_i\delta(s) Q\theta_\beta
    \right\}
    \right\} ,\Sigma_{\theta_{\gamma_j}(s)|\rest}\right)\\
    \delta(s)|\rest &= 1\times P_1(s) + 0\times P_0(s)\\
    P_1 & \propto \exp\left\{-\frac{1}{2\sigma_Y^2}\left[
    \delta^T\diag\br{\beta^2\|X_i\|_2^2}\delta 
    -2\mbr{\sum_i^nX_i(Y_i-\eta_i-\sum_{k=1}^{m}\gamma_j Z_{i,k})}^T\diag\br{\beta}\delta
    \right] \right\}
\end{align*}
Denote $D_{\delta}:= Q^T\diag\br{\delta}Q$. The log-Likelihood w.r.t. $\theta_\beta$ can be expressed as 
\begin{align*}
    \log L(\theta_\beta)&=-\frac{1}{2\sigma_Y^2}\sum_{i=1}^n \|Y_i^*-\theta_{\eta,i}^*(s) -X_i D_{\delta}\theta_\beta \|_2^2\\
    &=\frac{1}{\sigma_Y^2}\left(\sum_{i=1}^n (Y_i^*-\theta_{\eta,i}(s)- \sum_{k=1}^{m}\theta_{\gamma,k} Z_{i,k})X_i\right)^T D_\delta\theta_\beta - \frac{\sum_{i=1}^nX_i^2}{2\sigma_Y^2} \theta_\beta^TD_\delta^2\theta_\beta\numberthis\label{eq:logL}
\end{align*}
Based on the above derivations, we pre-compute the following summary statistics. These notations will help readers understand our code if one is interested.
\begin{itemize}
    \item \texttt{XY\_term\_allsample}: $\sum_{i=1}^nX_i Y_i(s) \in \R^p$
    \item \texttt{XqYstar\_term\_allsample}: $Y^*_{L\times n}Z_{n\times m}\in \R^{L\times m}$
    \item \texttt{XXq\_sumsq}: $\sum_{i=1}^nX_i Z_i\in \R^m$
    \item \texttt{XcXq\_sumsq}: $\sbr{\sum_{i=1}^n Z_{i,j} Z_{i,[-k]}}_{j=1}^q\in \R^{(m-1)\times m}$
    \item \texttt{XY\_eta\_term}: $\sum_{i=1}^n X_i Y_i(s) - \eta_{p\times n} X_{n\times 1} \in \R^{p\times 1}$
    \item \texttt{XqYstar\_theta\_eta\_term}: $Y^*_{L\times n} - \theta_{\eta,(L\times n)} \in \R^{L\times n}$
\end{itemize}

From the above derivation, we can see that the posterior covariance matrix for $\theta_\beta$ is a dense matrix, hence we only apply the SGLD algorithm on $\theta_\beta$.

\subsection{Imputing missing values using individual masks}
In this subsection, we consider how to update the missing outcome $Y_{i}(s_j)$ when 
considering the individual masks. Denote $Q_i$ to be the $i$-th rows in $Q$. 
For subject $i$ with missing voxel $s_j\notin \cV_i$,
\begin{align}\label{eq:Y_imp}
    Y_i(s_j) &= X_i \delta(s_j) Q_{j} \theta_\beta + \sum_{k=1}^q Q_{j} \theta_{\gamma,k} Z_{i,k} + Q_{j}\theta_{\eta,i} + \epsilon_i
\end{align}
In practice, to avoid repeated access of the entire data set, we save all locations of missing voxels in a vector \texttt{Y\_imp}. When accessing one batch of outcome data, the corresponding missing locations in $Y_i(s)$ is replaced by the imputed data saved in \texttt{Y\_imp}.

Below is the detail algorithm for the imputation part. We use $\cV_i^c$  to denote the collection of all missing voxels for subject $i$.

\begin{algorithm}[ht!]
\caption{Update missing outcomes at one iteration for subjects in the $b$ batch.}
\label{algo1_miss}
\begin{algorithmic}[1]
\State Let \texttt{Y\_imp} be the vector $\br{Y_i(s_j): s_{j}\in \cV_{i}^c, i=1,\dots, n, j=1,\dots,p}\in \R^{\sum_{i=1}^n|\cV_i^c|}$. The length of \texttt{Y\_imp} is the total number of missing outcomes across subjects and across locations.
\State Let $\cM$ be a list of the index map that stores the location of missing voxels in \texttt{Y\_imp}, i.e. the missing voxel $Y_i(s_j)$ is located at the $k$-th element in \texttt{Y\_imp}.
\For{ region $r=1,\dots R$,} 
    \State{Extract index set $\cV_{r}^c$, a list of index vectors, the $i$-th element is the vector $\cV_i^c \cap \cB_r$.}
    \State Extract index set $\cM_r$, a list of index vectors, the $i$-th element is the vector  $\cV_i^c \cap \cM$.
    \For{subject $i$ with in batch $b$ }
    \State Extract index vector $\cV_{r,i}^c$ which is the $i$th element in $\cV_{r}^c$
    \State Extract index vector $\cM_{r,i}$ which is the $i$th element in $\cM_r$
    \If{ $\cV_{r,i}^c$ and $\cM_{r,i}$ are not empty set}
    \State Update \texttt{Y\_imp} at locations $\cM_{r,i}$ using \eqref{eq:Y_imp} where $Y_i(s_j), s_j\in \cV_{r,i}^c$.
    \EndIf
    \EndFor
\EndFor
\end{algorithmic}
\end{algorithm}

Note that when computing the likelihood, we need the imputed values from \texttt{Y\_imp}. The reverse mapping from individual $i$ location $s_j$ to a particular $k$-th element in \texttt{Y\_imp} is similar to the above algorithm.

\section{Main Simulation Results } \label{sec:simulation}

\subsection{Simulation Design}\label{subsec:sim_design}

\textbf{Mass Univariate Analysis (MUA)} is one of the most commonly used methods for ISR models. MUA ignores any spatial correlation in the image and treats the data at all voxels as completely independent and performs linear regression on the data at each voxel. Ignoring spatial correlation may lead to potential overfitting and low power. When comparing selection accuracy for $\beta(s)\delta(s)$, we use the Benjamini-Hochberg adjusted p-values \citep{Benjamini1995-re} to control the false discovery rate. The missing voxels are directly imputed as 0 in the MUA. Because the volumetric fMRI data are preprocessed as t-statistics, 0 imputation is widely used in brain image ISR models since 0 acts as the null hypothesis with no significant neural activities.

\textbf{Bayesian Image-on-Scalar (BIOS) model} is the baseline Bayesian method for our proposed model \eqref{eq:model}-\eqref{eq:eta_prior}, where all parameters are sampled via Gibbs sampling. Missing voxels are imputed as 0. The sampling algorithm reads the entire data set into memory.

\textbf{Scalable Bayesian Image-on-Scalar model with 0 imputation (SBIOS0) } is our proposed model \eqref{eq:model}-\eqref{eq:eta_prior} with sampling performed via the SGLD algorithm and memory-mapping. Missing data at voxels in individual masks are replaced by 0.

\textbf{Scalable Bayesian Image-on-Scalar model with GS-imputation (SBIOSimp) } is the proposed model \eqref{eq:model}-\eqref{eq:eta_prior} using the SBIOS0 algorithm, however, missing data in the individual masks are imputed from the proposed model. 

\begin{figure}[ht]
\centering
\centering
  \begin{subfigure}{0.35\textwidth}
    \centering
    \includegraphics[width=\linewidth]{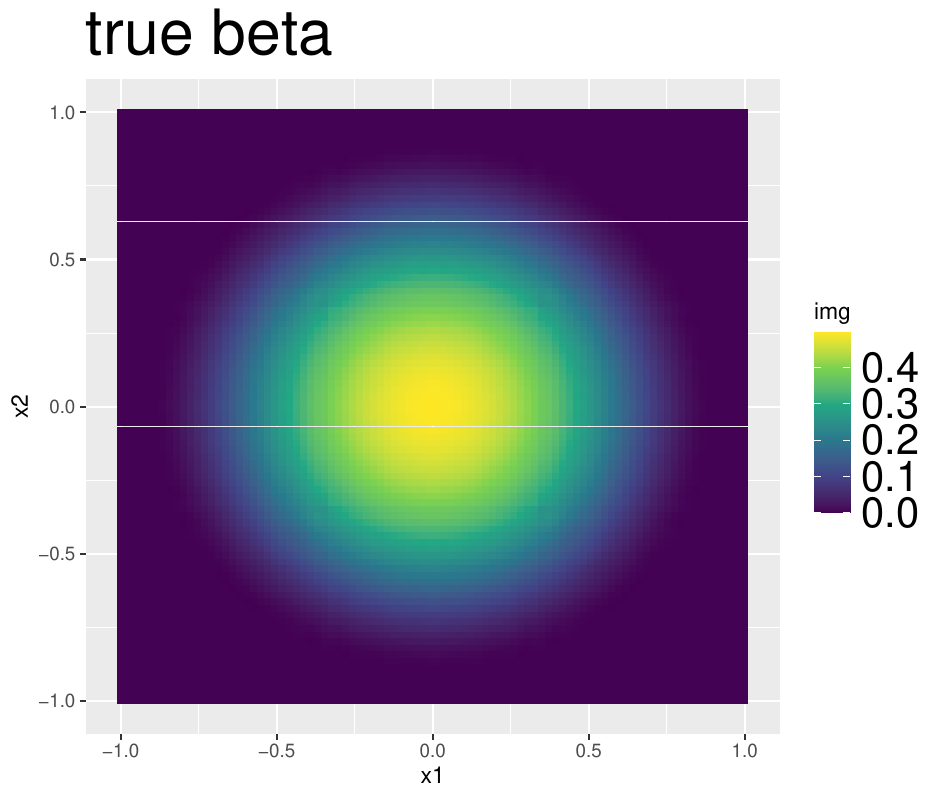}
    \caption{True covariate $\beta(s)\delta(s)$.}
    \label{fig:figure1}
  \end{subfigure}
  \hfill
  \begin{subfigure}{0.30\textwidth}
    \centering
    \includegraphics[width=\linewidth]{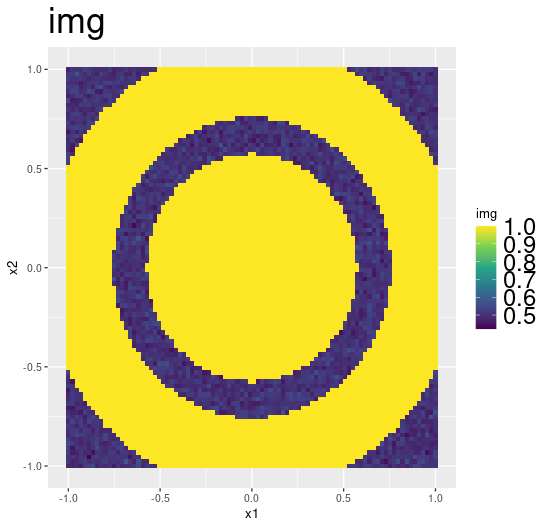}
    \caption{Missing Pattern I}
    \label{fig:true_beta_mis1}
  \end{subfigure}
  \begin{subfigure}{0.30\textwidth}
    \centering
    \includegraphics[width=\linewidth]{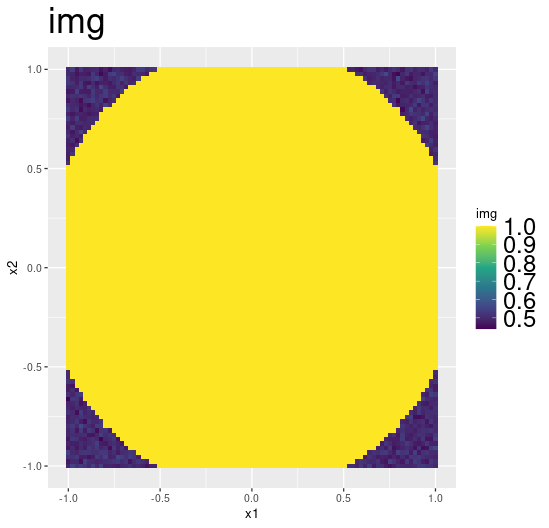}
    \caption{Missing Pattern II}
    \label{fig:true_beta_mis2}
  \end{subfigure}
\caption{True signal and missing patterns. Missing Pattern I \& II: OP ranges from 0.5 to 1.}
\label{fig:true_beta}
\end{figure}

To demonstrate the scalability of the proposed method, three simulation studies are used to show the selection accuracy (Simulation I), maximum memory usage (Simulation II), and time scalability (Simulation III). In each simulation, the entire sample is split into $B$ batches of data, each containing 500 subjects, with subsample size $n_s$ fixed at 200. The true image $\beta(s)$ is a 2D image of size $p=90\times 90=8{,}100$, as shown in Figure \ref{fig:true_beta} (a). The binary variable $\delta(s)$ is generated as 1 when $\beta(s)$ is nonzero, and 0 otherwise. The confounder parameters $\gamma_k(s)$ and individual effects $\eta_i(s)$ are generated based on the coefficients $\theta_{\gamma,k,l},\theta_{\eta,i,l}$, sampled from the standard normal distribution. Figure \ref{fig:true_beta_mis1} and \ref{fig:true_beta_mis2} are visual illustrations of the spatial observed proportion over all samples, where the image value at each pixel $s_j$ is $n^{-1}\sum_{i=1}^nI(s_j\in \cV_i)$. The difference between these 2 missing patterns is that missing pattern I selects part of the active area where $\beta(s)\neq 0$ with missing proportion of 50\% while missing pattern II only selects missing voxels outside of the active area.

When generating the covariance kernel function, the $90\times 90$ 2D images are evenly split into 9 ($3\times 3$) regions that are assumed to be independent from each other, a priori. Hence we compute the basis decomposition of each region supported on a square of $30\times 30$ pixels, where the kernel function is the Mat\'ern kernel with $\rho=2,\nu=1/5$ as in \eqref{eq:matern}. 

The number of basis coefficients used is 10\% of the number of pixels, hence we have 90 basis coefficients for each region. For a less smooth input signal pattern, we can use a larger number of basis functions to increase the flexibility of the fitted functional coefficients. 

All three Bayesian methods (BIOS, SBIOS0, SBIOSimp) use the same basis decomposition of the kernel function in \eqref{eq:matern}, and all run for a total of 5,000 MCMC iterations, where the last 1,000 iterations are used for posterior inference. The prior for $\delta(s)$ is Bernoulli$(0.5)$. For the SGLD based methods (SBIOS0, SBIOSimp), we tune the step size $\tau_t$ across iterations using the formula $\tau_t = a(b+t)^{-\gamma}$
as in \cite{Welling_undated-zi}, and set $a = 0.001, b=10,\gamma=0.55$ to ensure that through the 5000 MCMC iterations, the step size $s_t$ decreases from around $2\times 10^{-3}$ to $9\times 10^{-6}$. The initial values for all three Bayesian methods are identical, where $\delta(s)$, $\theta_{\beta,l}$ and $\theta_{\gamma,k,l}$ are set to 1, $\theta_{\eta,i,l}$ are set to 0, and all variance parameters $\sigma_Y,\sigma_\beta,\sigma_\gamma,\sigma_\eta$ are set to 1.

\subsection{Simulation Results}

We demonstrate the advantage of our proposed SBIOS algorithm concerning three objectives: variable selection accuracy, maximum memory usage, and scalability in terms of computing time as the sample size increases. The main results for selection accuracy, memory usage, and time scalability are shown in Figures \ref{fig:sim_TPR}, \ref{fig:mem_cost}, and \ref{fig:sim_all_time_bothN} respectively. We define the \textbf{observed proportion (OP)} at location $s_j$ to be $h(s_j) = n^{-1}\sum_{i=1}^n \indi_{\cV_i}(s_j)$, where $\cV_i$ is the set of all observed locations for individual $i$.

\subsubsection{Simulation I: Variable Selection}\label{subsec:sim1}

In Simulation I, to compare variable selection accuracy, when generating the individual masks, we provide results under two different missing data patterns as shown in Figure \ref{fig:true_beta} (b) and (c). Figure \ref{fig:true_beta} provides an illustration where in each pattern there exists a common area, and each location inside of this common area has observed data for all individuals. Outside of the common area, we allow missingness. This setting is motivated by the UK Biobank fMRI images where most missing voxels are at the brain periphery, and there is a large common area in the center of the brain with fully observed data. Outside of the commonly observed area, we allow the OP to be 0.5 or 0.1. Simulation I is performed under 8 scenarios: (i) OP of 0.5 or 0.1; (ii) $\sigma_Y = 0.5$ or $\sigma_Y=1$; (iii) missing pattern follows Figure \ref{fig:true_beta} (b) or (c). Changing the OP and missing patterns show the advantage of using imputation (SBIOSimp), and changing $\sigma_Y$ demonstrates algorithm performance under different signal-to-noise ratios.

\begin{figure}[ht!]
\centering
\includegraphics[width=0.6\textwidth]{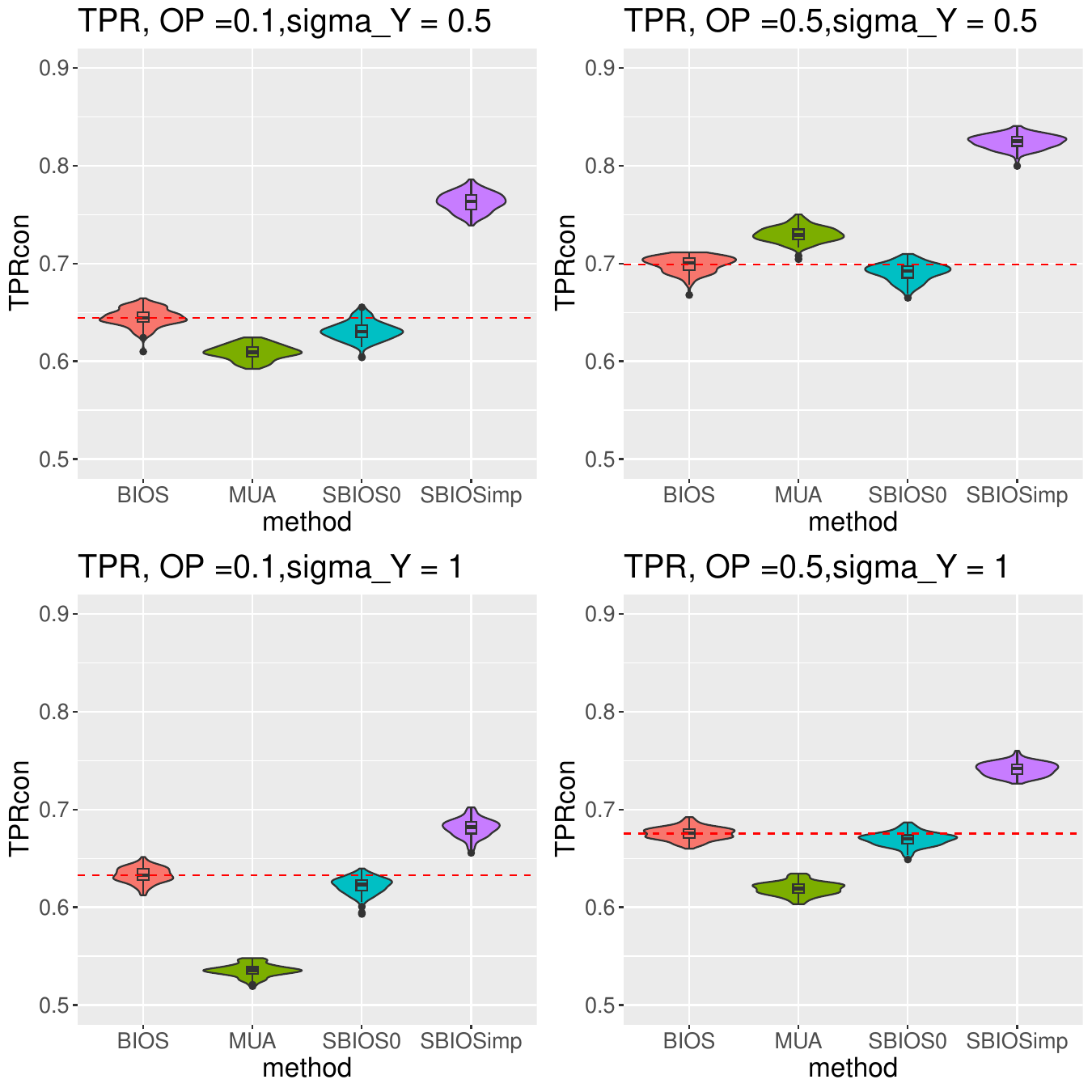}
\caption{True Positive Rate based on 100 replicated simulations when False Positive Rate is controlled at 0.1 when $n=3{,}000, p=8{,}100$.  Missing Pattern I.}
\label{fig:sim_TPR}
\end{figure}

The selection accuracy result in Figure \ref{fig:sim_TPR} shows that with Missing Pattern I, with the FPR controlled at 10\%, SBIOSimp has much better TPR across different settings compared to the other methods. Within the 0-imputation based methods, the Bayesian methods (BIOS, SBIOS0) have similar performances, and are generally better than MUA, except for the case where $\sigma_Y=0.5$, OP is 0.5. In situations where estimating $\beta\delta$ becomes challenging, such as when the signal-to-noise ratio is low or the OP is low, spatially correlated methods are advantageous. 

With Missing Pattern II, the three Bayesian methods outperform MUA in all settings. But since Missing Pattern II only generates missing pixels on non-active regions, there is smaller difference in FPR-controlled TPR between SBIOSimp and other methods, as shown in Section \ref{sec:supp_sim}. 

Comparing four sub-figures in Figure \ref{fig:sim_TPR}, we see that (i) as the OP decreases (comparing right subplots to left subplots), the TPR decreases for all methods, with MUA the most sensitive to the change in missing percentage; (ii) as the signal-to-noise ratio decreases ($\sigma_Y$ goes from 0.5 to 1, comparing top subplots to bottom subplots), the TPR decreases, again MUA being the most sensitive.

\begin{figure}[ht!]
        \centering
        \includegraphics[width=\linewidth]{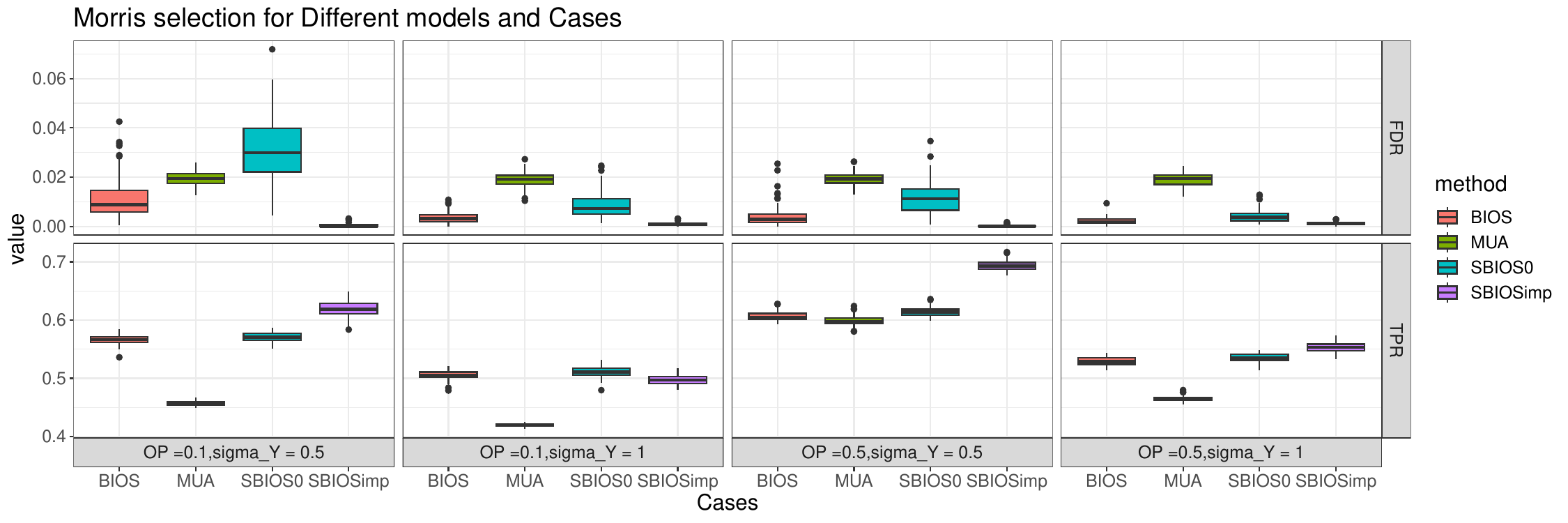}
        \caption{Use 95\% cutoff on PIP for Bayesian active voxel selection. }
        \label{fig:sim_PIP_main}
\end{figure}

Although Figure \ref{fig:sim_TPR} provides a fair comparison of the power of each method given the same FPR tuned according to the true signal pattern, in practice, we prefer criteria for active voxel selection that can control the FDR well without knowing the true signal. When we use the 95\% cutoff on the PIP for variable selection, the FDR and TPR are shown in Figure \ref{fig:sim_PIP_main}. This indicates that the 95\% cutoff on PIP allows us to control FDR reasonably well below 0.05. The Supplementary Figure \ref{fig:sim_3criteria} provides a full comparison among three selection criteria: (i) Morris' FDR control method \citep{Meyer2015-mf}; (ii) MUA-based selection, i.e. choosing the cutoff on PIP such that the Bayesian method would have the same proportion of active voxels as identified by MUA BH-adjusted p-values; (iii) Directly applying the 95\% cutoff on the PIP. The result in Figure \ref{fig:sim_3criteria} demonstrates that criteria (ii) and (iii) can both control the FDR well below 0.05, but the 95\% threshold gives a  better TPR for the Bayesian methods, hence we use this criterion as the cutoff on PIP in real data analysis.


\subsubsection{Simulation II: Memory Usage}
In simulation II, to compare memory usage, the generative setting sets $\sigma_Y =1$, missing pattern II, and an observed proportion of 0.5. The sample sizes are $n = 3000, 6000, 9000, 12000$, with fixed $p=3600$. SBIOSimp has similar memory usage to SBIOS0 except for a vector to save the imputed outcome, so we only compare MUA, BIOS, and SBIOS0.

To record maximum memory usage, we use the R function \texttt{Rprof}. In particular, we use \texttt{summaryRprof} with the option \texttt{memory="tseries"} where memory usage is recorded for segments of time. We take the maximum memory across all such segments as the maximum memory usage. The reported memory usage is for running each algorithm and does not include memory usage for reading data into memory. Reading memory is recorded separately from running memory. Detailed results can be found in Section \ref{sec:supp_sim}. For MUA and BIOS, we directly read all data into memory first, and pass the address to BIOS. For SBIOS0 and SBIOSimp, each batch of data are read into R as file-backed matrices on the disk, and its address is passed to both SBIOS0 and SBIOSimp for sampling.

\begin{figure}[ht]
\centering
\includegraphics[width=0.7\textwidth]{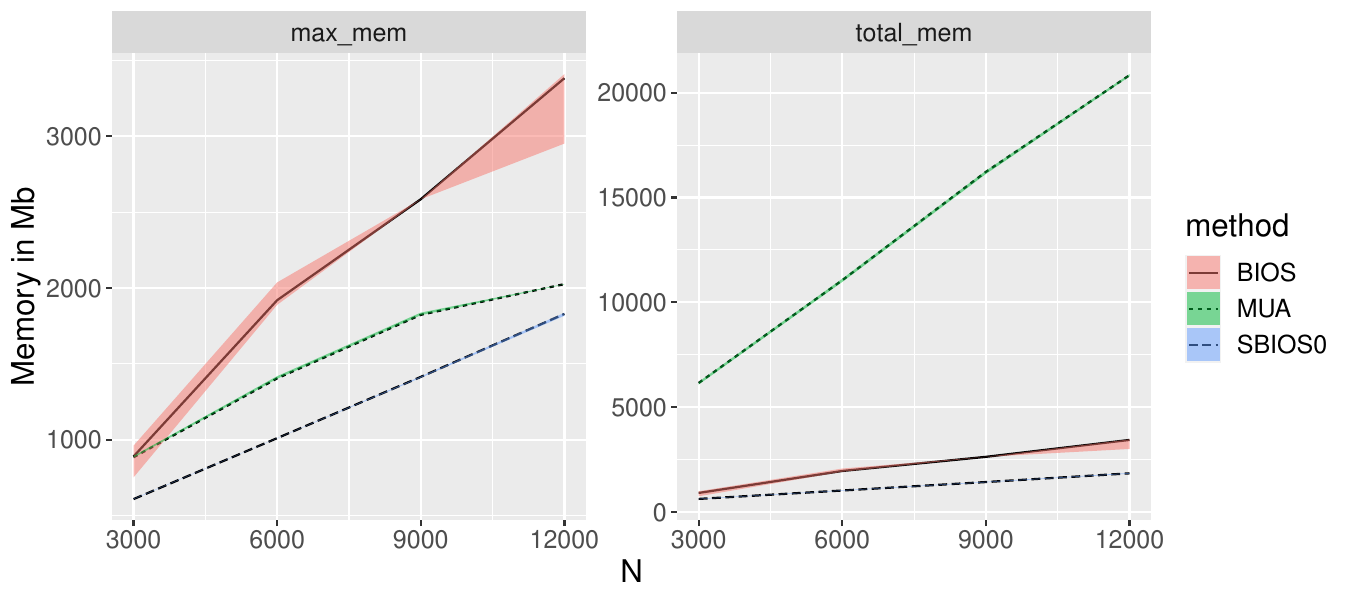}
\caption{Memory cost in Mb with $n = 3000, 6000, 9000, 12000$, using 100 replications. The left panel shows maximum memory used, and the right panel shows total memory used. The mean, 97.5\% and 2.5\% quantile lines are also displayed.}
\label{fig:mem_cost}
\end{figure}

Results are shown in  Figure \ref{fig:mem_cost}. The left panel in Figure \ref{fig:mem_cost} reflects the peak memory for each method, and is the main result of interest. In terms of maximum memory used over time, our proposed method SBIOS0 is comparable to MUA, which only performs linear regressions and is much more efficient than the standard Bayesian method, BIOS. In terms of total memory used over time (right panel in Figure \ref{fig:mem_cost}), both Bayesian methods take much less total memory than MUA which needs to loop through $p$ locations in total, but our proposed method SBIOS0 still outperforms the other.

\subsubsection{Simulation III: Time Scalability}
In simulation III, we compare time scalability for BIOS, SBIOS0, and SBIOSimp, and exclude MUA since it only needs to run $p$ simple linear regressions and thus is linear in the number of voxels. For this simulation,  the generative setting uses missing pattern II, and sample sizes $n = 3000$ and $6000$, with a fixed number of voxels at $p=3600$. 
\begin{figure}[!ht]
\centering
    \begin{subfigure}{\textwidth}
        \centering
        \includegraphics[width=0.8\linewidth]{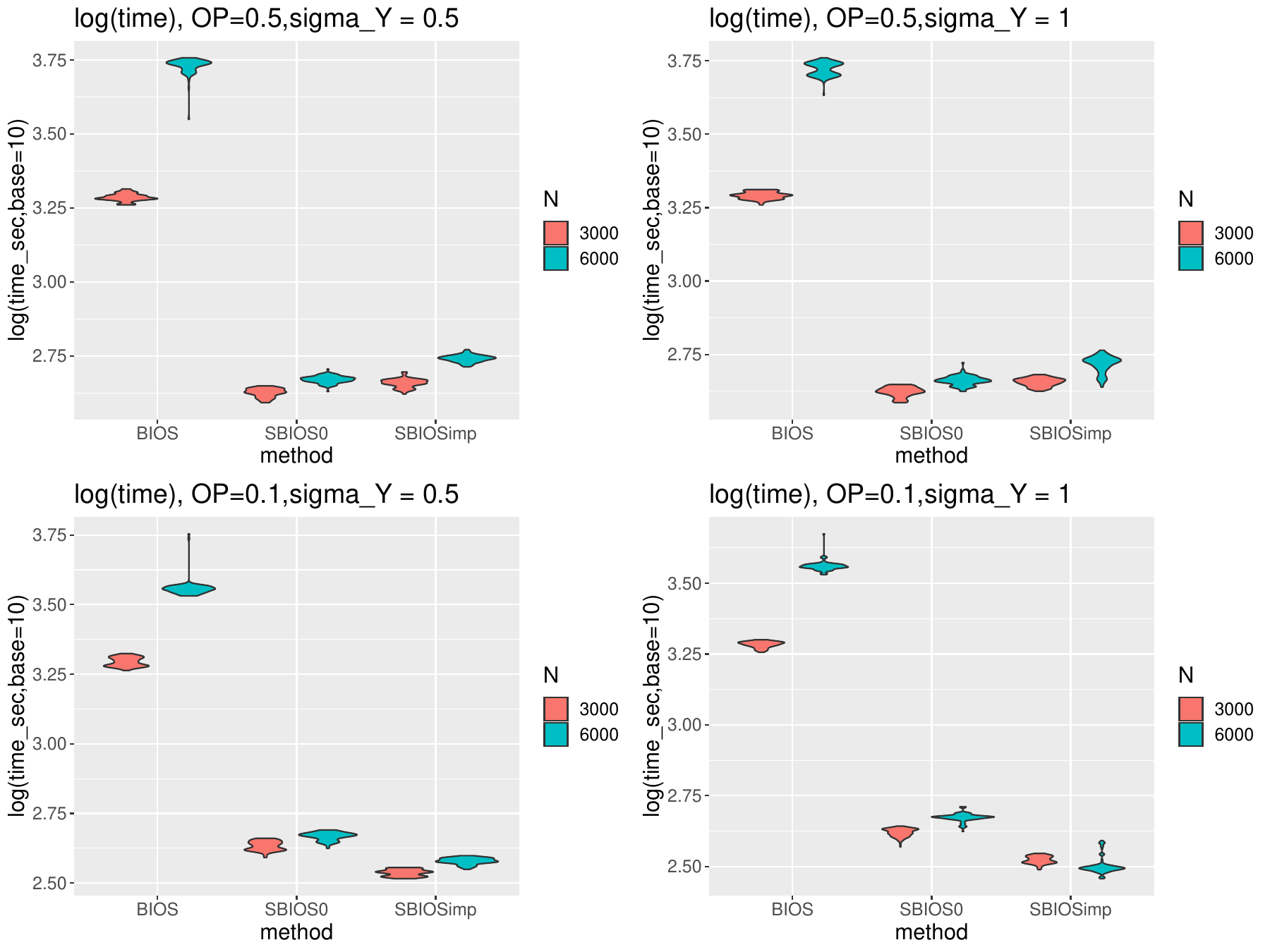}
        \caption{Computing time for Bayesian methods, in $\log_{10}(\text{seconds})$, for both $n=3000$ and $n=6000$ based on 100 simulations. }
        \label{fig:sim_all_time_bothN}
    \end{subfigure}
    \begin{subfigure}{\textwidth}
        \centering
        \begin{tabular}{lllll}
        \toprule
        method                    & \multicolumn{2}{c}{SBIOS0} & \multicolumn{2}{c}{SBIOSimp} \\\midrule
        $n$                         & 3000        & 6000         & 3000         & 6000          \\
        OP=0.5, $\sigma_Y$ = 0.5 & 4.57        & 11.50        & 4.24         & 9.77          \\
        OP=0.1, $\sigma_Y$ = 0.5 & 4.58        & 7.78         & 5.73         & 9.56          \\
        OP=0.5, $\sigma_Y$ = 1   & 4.66        & 11.53        & 4.32         & 10.11         \\
        OP=0.1, $\sigma_Y$ = 1   & 4.61        & 7.72         & 5.76         & 11.44        \\
        \bottomrule
        \end{tabular}
        \caption{Ratio of speed for SBIOS0 and SBIOSimp when compared to BIOS, i.e. ~time of BIOS over time of SBIOS0 (or SBIOSimp) under 4 cases and n = 3000 and 6000, based on 100 replicated simulation results.}
        \label{tb:sim_all_time_bothN}
    \end{subfigure}
    \caption{Computational time comparison}
\end{figure}


As illustrated in Figure \ref{fig:sim_all_time_bothN}, the computing time (seconds) is reported on the log base 10 scale. Our proposed SBIOS0 and SBIOSimp are shown to have much better time scalability across all settings compared to BIOS which uses the full data and Gibbs sampling.

To test the performance of SBIOSimp on the UK Biobank data, we have included two additional simulation studies in the supplementary material: Supplementary Section \ref{sec:supp_RDA_sim} provides a simulation with the same scale as UK Biobank data, where the true signal is simulated as a piecewise smooth function across different regions; Section \ref{sec:supp_oneKer} provides a low-resolution simulation using SBIOSimp but with a single GP kernel, instead of a region-wise independent kernel. Both simulation examples show that even when the kernel is misspecified and differs from the smoothness of the true signal, SBIOSimp can still identify the active voxels with good accuracy.

\section{Additional Simulation Results}\label{sec:supp_sim}
\begin{figure}[ht!]
\centering
\includegraphics[width=0.7\textwidth]{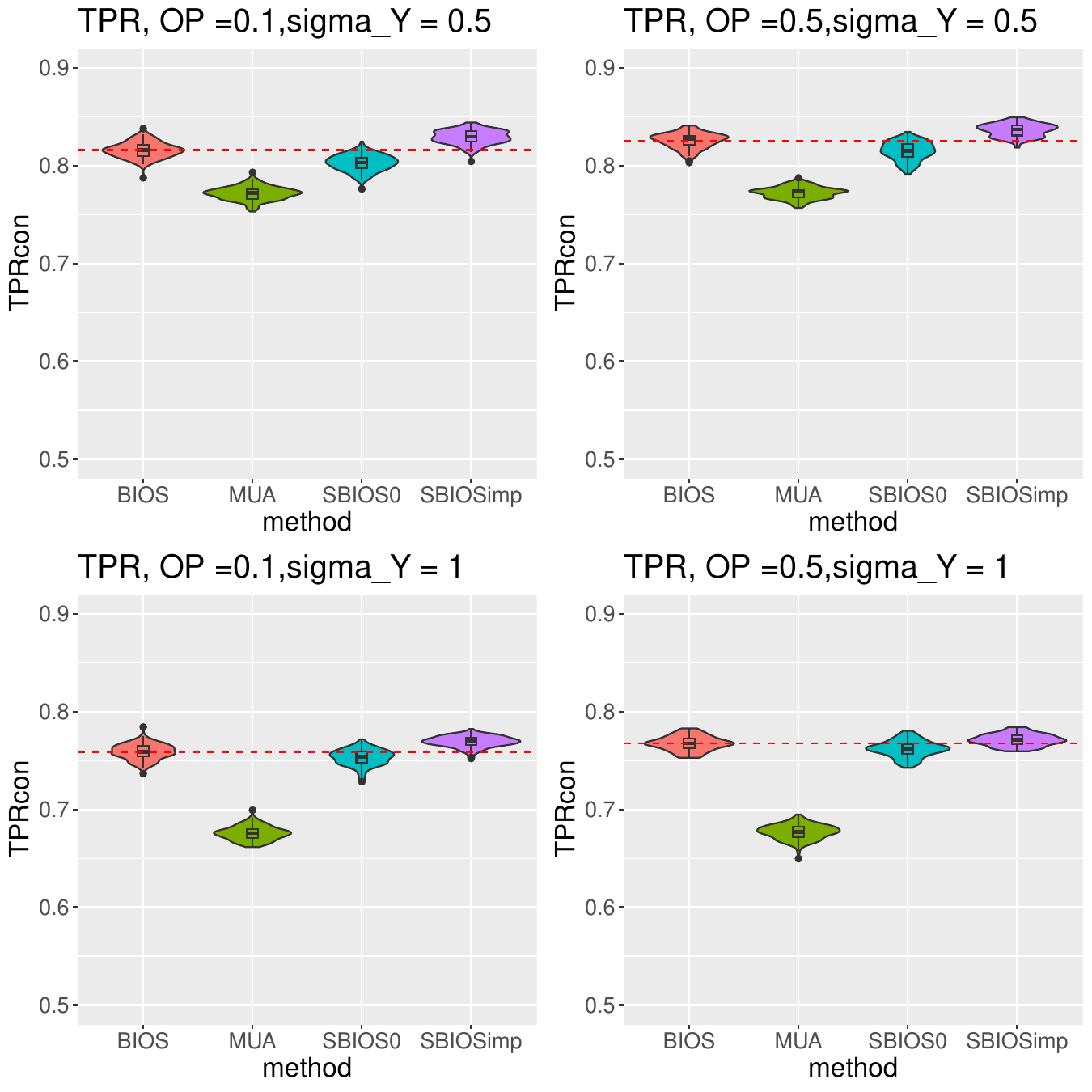}
\caption{True Positive Rate based on 100 replicated simulation results when False Positive Rate is controlled at 0.1 when $n=6000, p=8100$.  Missing Pattern II.}
\label{fig:sim_FDR_all_TPRcon}
\end{figure}

\begin{figure}[ht!]
\centering
\includegraphics[width=\textwidth]{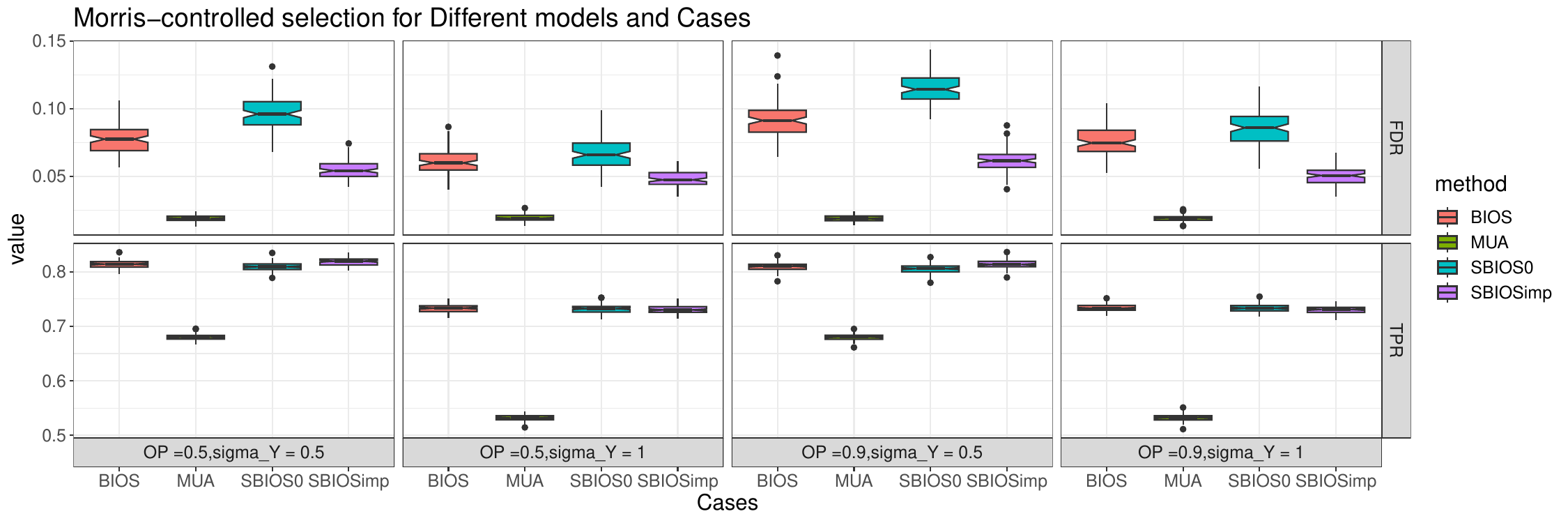}
\caption{True Positive Rate and False Discovery Rate using Morris FDR control method, based on 100 replicated simulations, $n=3000, p=8100$. Missing Pattern II.}
\label{fig:sim_Morris_misII}
\end{figure}

\begin{figure}[ht!]
\centering
\includegraphics[width=\textwidth]{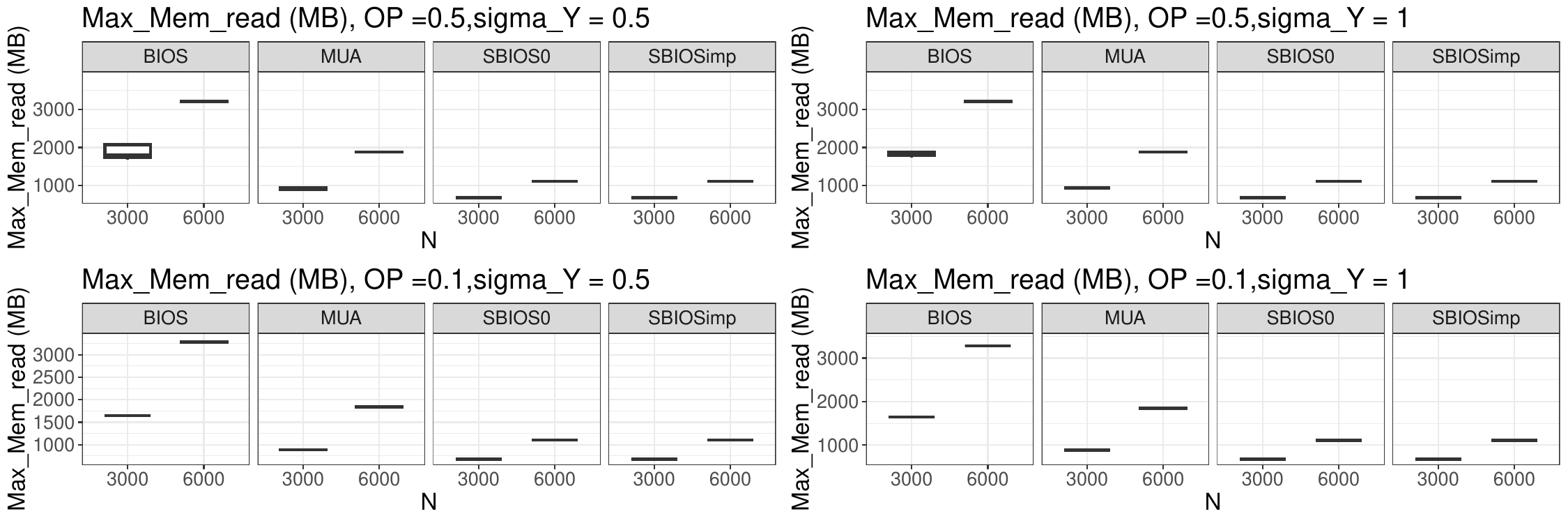}
\caption{Maximum total reading memory (MB) for both $n=3000$ and $n=6000$ based on 100 replicated simulation results.}
\label{fig:sim_all_max_mem_read_bothN}
\end{figure}

\begin{figure}[ht!]
\centering
    \begin{subfigure}{\textwidth}
        \centering
        \includegraphics[width=\linewidth]{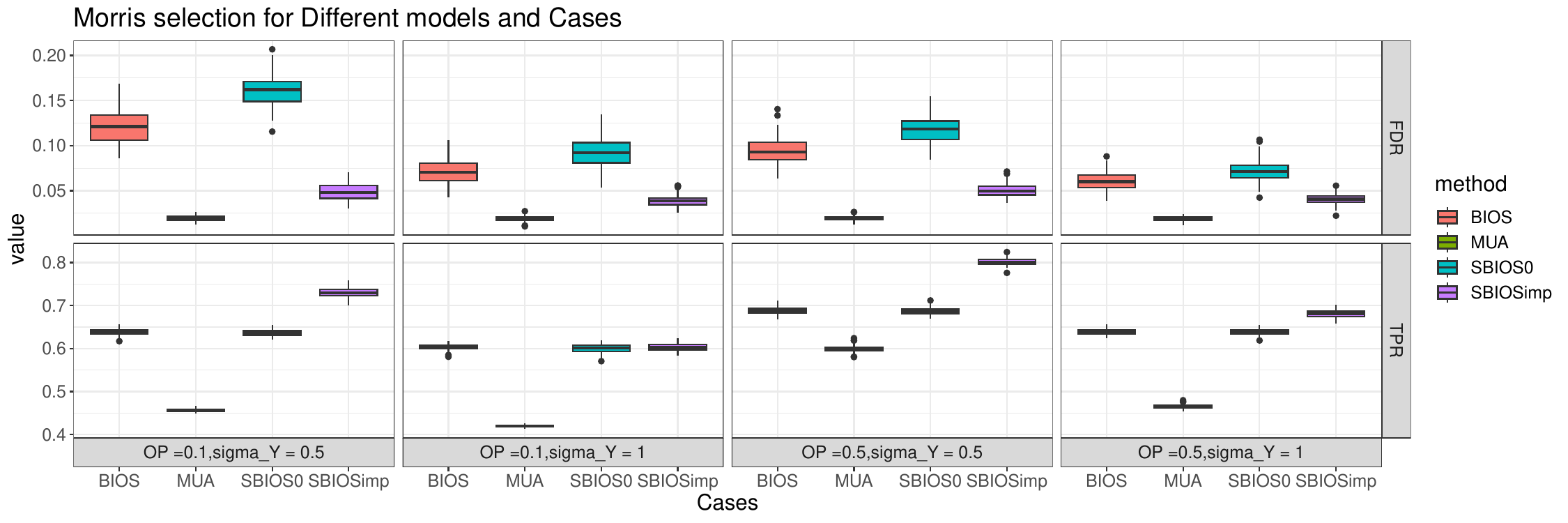}
        \caption{Morris' FDR control method}
        \label{fig:sim_Morris}
    \end{subfigure}
    \begin{subfigure}{\textwidth}
        \centering
        \includegraphics[width=\linewidth]{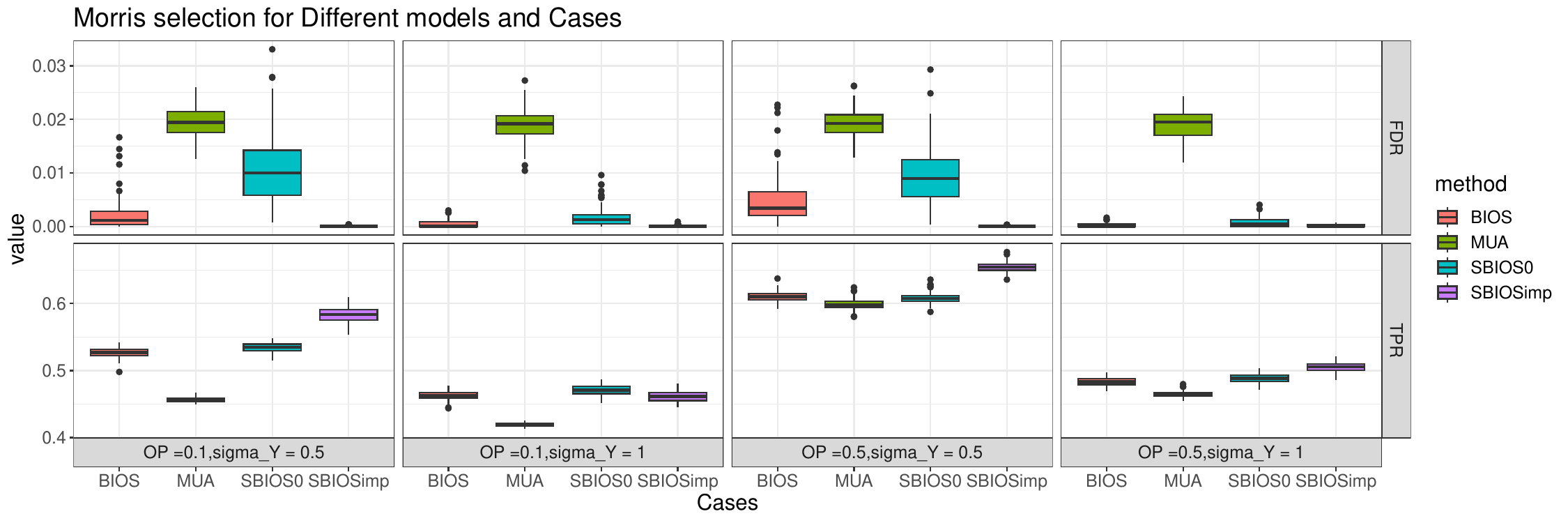}
        \caption{Use MUA's proportion of active voxels to choose the cutoff on PIP.}
        \label{fig:sim_MUA}
    \end{subfigure}
    \begin{subfigure}{\textwidth}
        \centering
        \includegraphics[width=\linewidth]{plot/sim_newmis_2criteria_all_PIP_selection_new.pdf}
        \caption{Use 95\% cutoff on PIP.}
        \label{fig:sim_PIP}
    \end{subfigure}
\caption{True Positive Rates (TPR) and False Discovery Rates (FDR) using three criteria to choose the cutoff on PIP, based on 100 replicated simulations, $n=3000, p=8100$. Missing Pattern I.}
\label{fig:sim_3criteria}
\end{figure}

This section provides additional simulation results, including the TPR result for Missing Pattern II (Figure \ref{fig:sim_FDR_all_TPRcon}), the Morris FDR controlled selection results for  Missing Pattern II (Figure \ref{fig:sim_Morris_misII}), the reading memory cost (Figure \ref{fig:sim_all_max_mem_read_bothN}), and the comparison among three active voxel selection criteria (Figure \ref{fig:sim_3criteria}).

Below, Figure \ref{fig:sim_FDR_all_TPRcon} provides the simulation results in Simulation I with missing Pattern II. The TPR is generally greater than with missing Pattern I since the missing pixels in missing Pattern II are all located outside of the active region. MUA presents the worst performance and SBIOSimp has the best performance across all settings, but the difference between SBIOSimp and the other competing methods is much smaller compared to the result in Simulation I with missing Pattern I.

Figure \ref{fig:sim_Morris_misII} provides the simulation results in Simulation I with missing Pattern II when applying Morris FDR control method to the Bayesian methods.

Figure \ref{fig:sim_all_max_mem_read_bothN} provides additional results on the reading memory in Simulation II: Memory Usage. Here, both SBIOS0 and SBIOSimp read $B$ batches of data in loops, and save them as file-backed matrices, hence the memory usage increase for reading memory is smaller compared to BIOS and MUA.

Figure \ref{fig:sim_3criteria} provides the comparison of FDR and TPR among three criteria: (i) Morris FDR control method \citep{Morris2015-aj}; (ii) selection based on the MUA's proportion of active voxels (determined by the BH-adjusted p-values less than 0.05); (iii) 95\% cutoff on the PIP. Based on Figure \ref{fig:sim_3criteria}, we can see that the Morris FDR control method cannot control the FDR below the target (0.05). The proportion of MUA can control the FDR well below 0.05 for the Bayesian methods, but the power is lower than the 95\% cutoff on PIP.

\newpage

\section{Additional UKBiobank Results and Figures}\label{sec:supp_RDA}
\begin{figure}[ht!]
\centering
\includegraphics[width=0.9\textwidth]{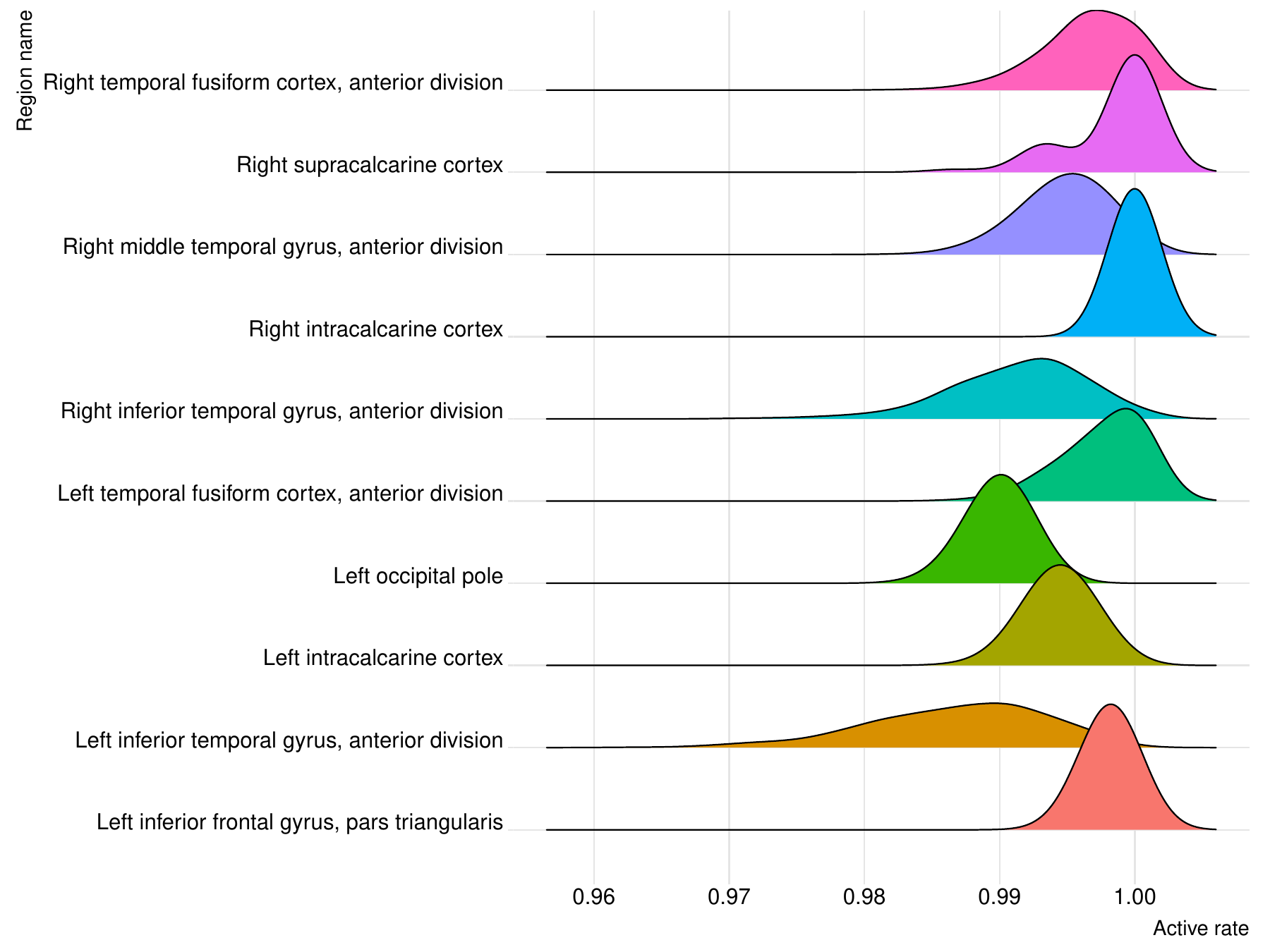}
\caption{Posterior distribution of the region level active rate (RLAR) for the top 10 regions with highest mean region active rate.}
\label{fig:RDA_top10region}
\end{figure}

\begin{figure}
\begin{tabular}{cc}
    \centering
    \includegraphics[width=0.5\textwidth]{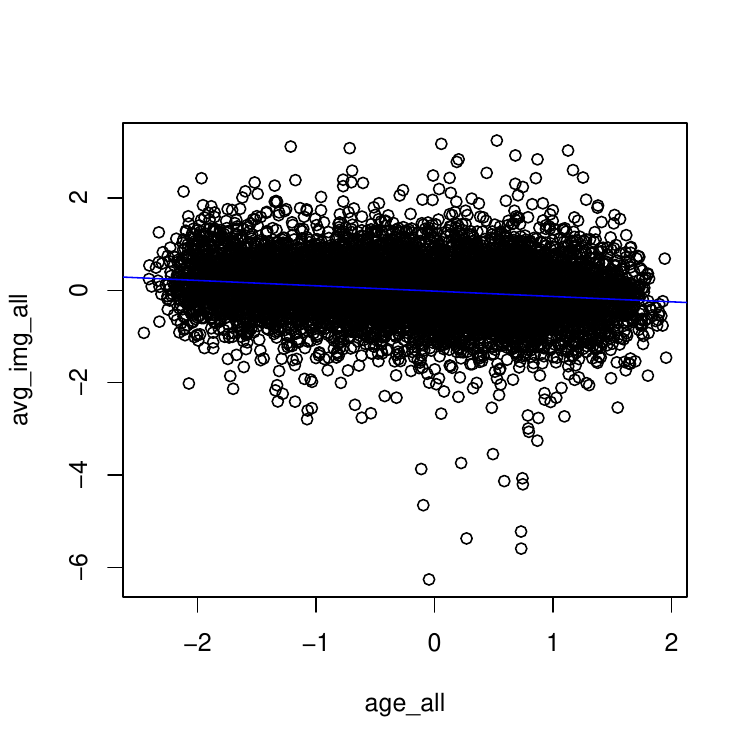} &   \includegraphics[width=0.5\textwidth]{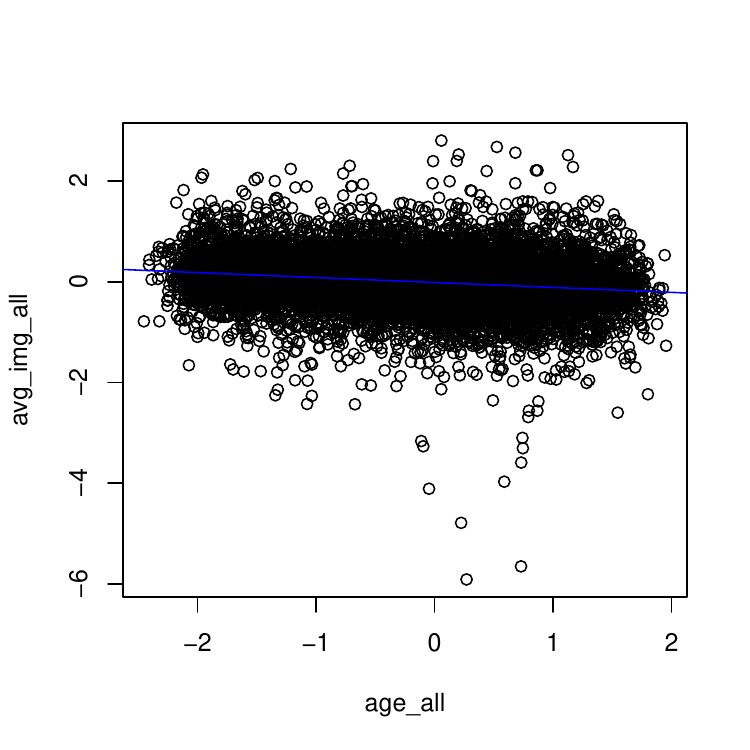} \\
    SBIOS0 & SBIOSimp\\
    \begin{tabular}{rrrr}
\hline
est. & Lower CI & Higher CI & p.value\\
-0.1731 & -0.1938 & -0.1523 & $<0.001$\\
\hline
\end{tabular} & \begin{tabular}{rrrr}
\hline
est. & Lower CI & Higher CI & p.value\\
-0.1666 & -0.1873 & -0.1457 & $<0.001$\\
\hline
\end{tabular}
    \end{tabular}
    \caption{Scatter plot of average image intensity(standardized) over the largest connected area against  Age(standardized) for all individuals. The actual age ranges from 44 to 83. The connected area is defined as follows: first we apply 0.7 cutoff on the inclusion probability to select active voxels with IP$>$0.7, then we search for the largest area with most connected active voxels. The scatter plot on the left is based on the inclusion probability obtained from SBIOS0 method, which happens to be a subset of the largest connected area obtained based on SBIOSimp. The right plot is based on GS-imputation.}
    \label{fig:age_plot}
\end{figure}

 Figure \ref{fig:RDA_top10region} is a histogram of the RLAR posterior distribution for the top 10 identified regions. Figure \ref{fig:age_plot} provides the scatter plot based on the observed data, indicating that there exists a negative association between age and brain signal intensity. Note that in Figure \ref{fig:age_plot}, we compute the averaged image intensity over the largest connected active areas in order to downplay the vast majority of brain voxels with no signals. If we average over the entire brain and fit a simple linear regression with the standardized age, the linear effect of age is $-0.041$ with p-value $0.02$, which is very close to 0.

\subsection{Calculation of percentage decline in Table 1}\label{supp_sec:percentage}

In Table 1, we include the percentage decline in the brain signal intensity associated with 10-year increase in age from 50 to 60, and 60 to 70, respectively. Here is the mathematical definition and derivation for this percentage. This derivation is based on the assumption that age is independent of the confounders gender and headsize. This is verified in the observed data where the empirical correlation between age and gender is 0.084, between age and headsize is -0.04, both close to 0.

We denote $A_i$ as age, $C_{1,i}$ as gender, $C_{2,i}$ as headsize. The confounder coefficient for the interaction of age and gender is denoted as $\tilde\beta$. Let age increase from $a_0$ to $a_1$. 
\begin{align*}
    M_i(s) &= \beta(s)A_i + \tilde\beta(s)A_iC_{1,i} + \gamma_1(s)C_{1,i} + \gamma_2(s)C_{2,i} + \eta_i(s)+\epsilon_i(s)\\
    \bE\br{M(s)|A=a_0} &= \bE_{C} \bE\br{ M_i(s)|A=a_0,C } = \int  \bE\br{ M_i(s)|A=a_0,C } \dd \hat \bP(C|A=a_0)
\end{align*}
Let $\hat \bP(C|A=a_0)$ be the conditional empirical distribution of $C$ given $A$, where $C$ is the collection of $C_1,C_2$.

\begin{align*}
    \bE\br{M(s)|A=a_0}&=\beta(s)a_0 + \tilde\beta(s)a_0\bar C_1 + \gamma_1(s)\bar C_1 + \gamma_2(s)\bar C_2 + o_p(1)
\end{align*}
Here, $\bar C_1, \bar C_2$ is the sample mean of the confounders. The individual effects $\eta_i(s)$ and noise $\epsilon_i(s)$ are both assumed to be centered at 0 over $i$.

The percentage used in Table 1 is defined as follows.
\begin{align*}
    &\frac{\bE\br{M(s)|A=a_1} - \bE\br{M(s)|A=a_0}}{\bE\br{M(s)|A=a_0}} \times 100\%\\
    &\approx
    \frac{\beta(s)a_1 + \tilde\beta(s)a_1\bar C_1 - \beta(s)a_0 - \tilde\beta(s)a_0\bar C_1}{\beta(s)a_0 + \tilde\beta(s)a_0\bar C_1 + \gamma_1(s)\bar C_1 + \gamma_2(s)\bar C_2}\times 100\%\\
    &= \frac{\beta(s)\Delta a + \tilde\beta(s)\Delta a\bar C_1}{\beta(s)a_0 + \tilde\beta(s)a_0\bar C_1 + \gamma_1(s)\bar C_1 + \gamma_2(s)\bar C_2}\times 100\%
\end{align*}

\subsection{Empirical Covariance v.s. GP kernel}\label{supp:subsec:cov}

The smoothness parameters of the GP kernel used in the main analysis are chosen to roughly match the empirical covariance. We include a few scatter plots of the empirical covariance estimated using the observed image data v.s. the covariance created by the GP kernels, on a few selected regions. We also include this comparison on the posterior level of $\eta_i$ on a few pairs of spatially connected regions to check the impact of the region-specific GP prior.

\begin{figure}[ht]
  \centering

  \begin{subfigure}[t]{0.9\linewidth}
    \centering
    \includegraphics[width=\linewidth]{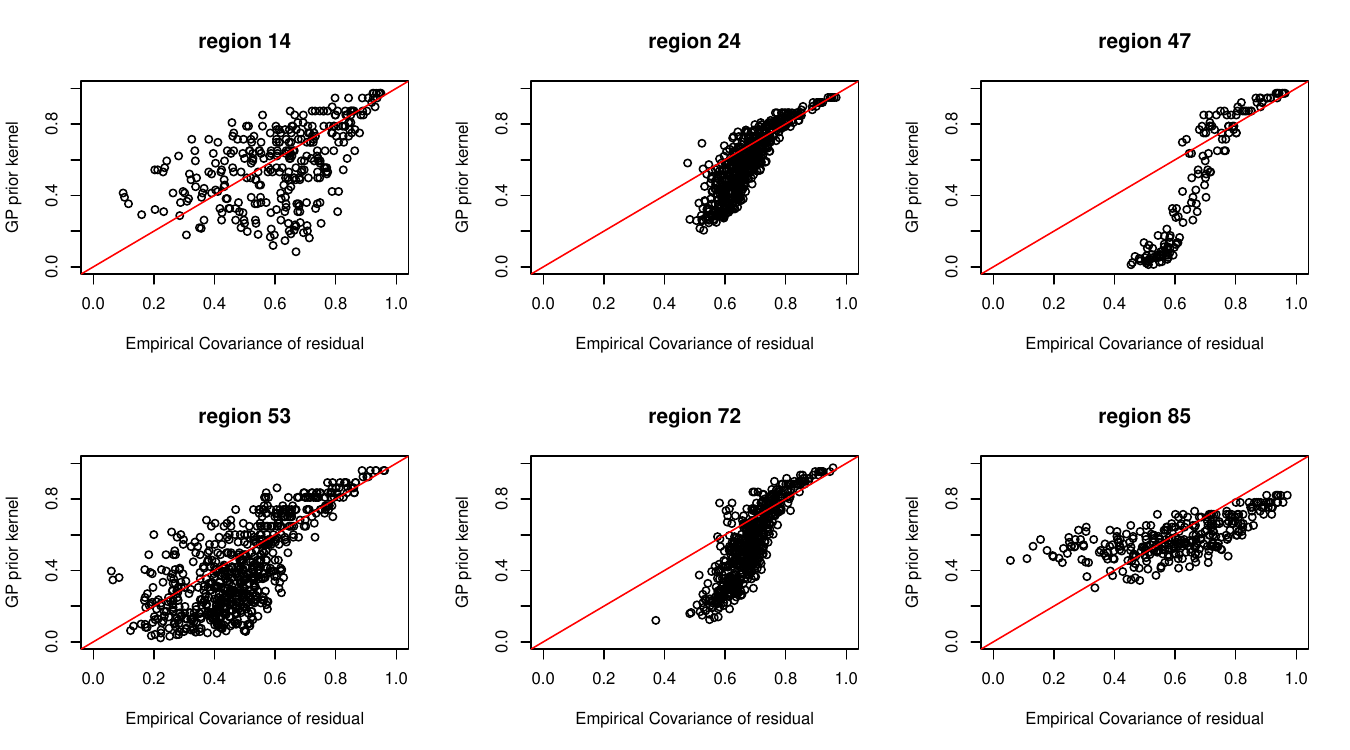}
    \caption{Within-region correlation (prior-level): Scatter plot of the GP prior kernel v.s. the empirical covariance of the standardized residual image $Y_i(s)-\hat\beta(s)X_i-\sum_{k=1}^m\hat\gamma_k(s_j)Z_{ik}$, on the top 6 regions with a large amount of active voxels as reported in Table~3.}
    \label{fig:Single_region_prior}
  \end{subfigure}

  \vspace{0.6em} 

  \begin{subfigure}[t]{0.9\linewidth}
    \centering
    \includegraphics[width=\linewidth]{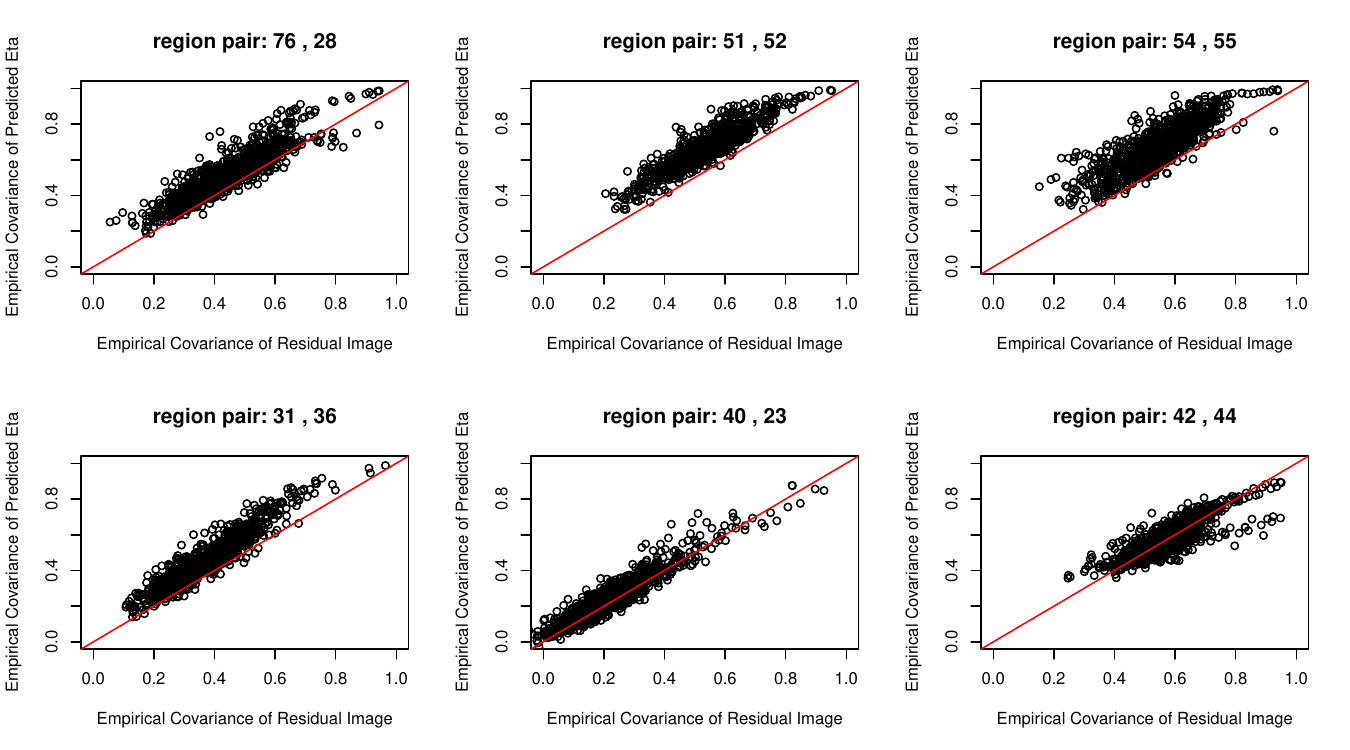}
    \caption{Cross-region correlation (posterior-level): Scatter plot of the empirical covariance of the residual image $Y_i(s)-\hat\beta(s)X_i-\sum_{k=1}^m\hat\gamma_k(s_j)Z_{ik}$ v.s. the empirical covariance of $\eta_i(s)$ directly drawn from the posterior full conditional given $\hat\beta(s),\hat\gamma_k(s_j),\hat\sigma_Y^2$, on 6 pairs of regions. The two regions in every pair are spatially connected.}
    \label{fig:Cross_region_eta}
  \end{subfigure}

  \caption{Comparison of the prior GP kernel with the empirical covariance.}
  \label{fig:emp_cov}
\end{figure}

When choosing the kernel smoothness parameters, we used slightly more voxels ($p=122{,}690$) than the real data in our main analysis, on a subsample $n=950$ of the whole data. Figure~\ref{fig:Single_region_prior} provides the scatter plot on the prior level of the GP kernel v.s. the estimated empirical covariance of the residual image $Y_i(s)-\hat\beta(s)X_i-\sum_{k=1}^m\hat\gamma_k(s_j)Z_{ik}$, where $\hat\beta(s),\hat\gamma_k(s_j)$ are the posterior means. The residual images are standardized, and the prior variance of the GP kernel is set to 1, so the plots represent the correlation comparison. We randomly select 1000 pairs of voxels within each region to compute their pairwise empirical covariances. As shown in Figure~\ref{fig:Single_region_prior}, when choosing the kernel smoothness parameters, they are chosen to roughly match the GP kernel on the 45-degree red line with the empirical covariance of the observed image data.

The above correlation comparison is on the prior level. We also compare the empirical covariance matrix of the posterior $\eta_i(s)$ v.s. the residual images. In this comparison, we use the same data as our main analysis, where $p=121{,}865$, and we compute the empirical covariances based on the first batch of data, $n=768$. The posterior sample of $\eta_i(s)$ is not saved during training due to memory limit. Hence, we directly sample the posterior $\eta_i(s)$ only on the chosen voxels in these selected regions and on this small subset of data, and the posterior full conditional of $\eta_i(s)$ is computed given the block-diagonal GP prior and the likelihood conditional on the converged posterior means of all other parameters. Figure~\ref{fig:Cross_region_eta} is the comparison of the cross-region correlation. Each pair of regions in Figure~\ref{fig:Cross_region_eta} are spatially connected. We can see that although the prior of $\eta_i(s)$ is block-diagonal, meaning that for pairs of voxels in different regions, their prior correlation is 0, but the posterior of $\eta_i(s)$ roughly matches the correlation pattern of the residual images.

\section{UKB-scale Simulation Study}\label{sec:supp_RDA_sim}

We provide an additional simulation study in this section, where the dimension of the simulated image ($p=121,865$) is the same as the UKB data, and the number of simulated subject is $n = 10,000$, evenly split into 10 batches of $1,000$ subjects each batch. 

This simulation aims to check the potential impact of the prior region-wise independence assumption on the posterior estimation. Because we assume region-wise independence for GP kernels at the prior level, although theoretically, the posterior of $\beta$ will converge to the true spatially-varying function $\beta_0$, regardless of the prior independence structure, it adds validity to our analysis to empirically verify that the region-wise independence assumption at prior-level, given the scale of the UKB data, would not create discontinuous jumps across region boundaries.

\begin{figure}[ht!]
\centering
    \begin{subfigure}{\textwidth}
        \centering
        \includegraphics[width=\linewidth]{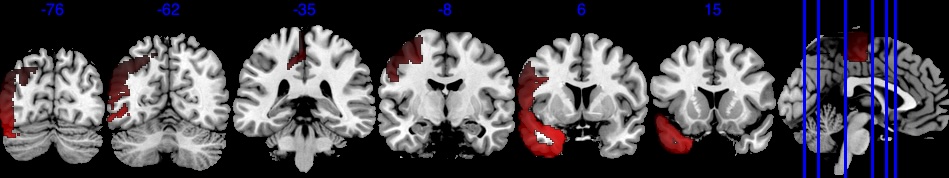}
        \caption{True $\beta(s)$ used in the simulated data, where $\beta(s)$ is generated as a smooth function across 4 regions 55,56,70,71.}
        \label{fig:UKBsim_beta_true}
    \end{subfigure}
    \begin{subfigure}{\textwidth}
        \centering
        \includegraphics[width=\linewidth]{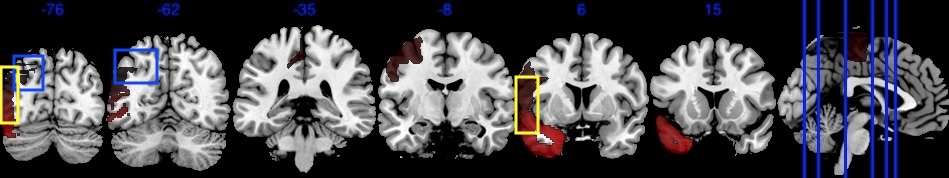}
        \caption{Estimated $\beta(s)$ by SBIOSimp. False negatives are highlighted with blue boxes. Region boundaries are highlighted with yellow boxes}
        \label{fig:UKBsim_beta_est}
    \end{subfigure}
    \begin{subfigure}{\textwidth}
        \centering
        \includegraphics[width=\linewidth]{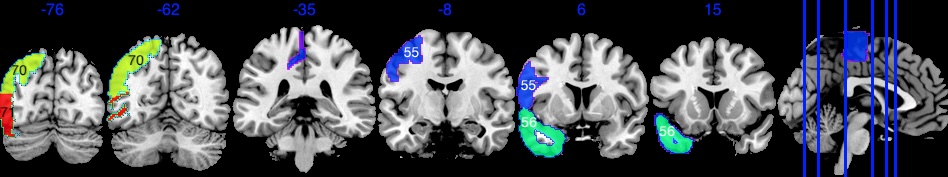}
        \caption{Region 55,56,70,71 for the parcellation used in the GP kernel.}
        \label{fig:UKBsim_beta_region}
    \end{subfigure}
\caption{True and estimated $\beta$ in the simulated data.}
\label{fig:UKBsim}
\end{figure}

The true $\beta$ is generated in Figure \ref{fig:UKBsim_beta_true}, a smooth function over pre-specified regions 55,56,70,71. Here, $\beta$ is simulated as $\beta(x) = \exp\{-0.01\|x-x_0\|_2^2/(10d)\}$, where $d=3$ is the dimension of $x$, and $x_0$ is the geometric center of all voxels in regions 55,56,70,71. Any values in $\beta(x)$ below 0.1 are thresholded to 0, so that the active area is contained in regions 55,56,70,71, but does not span over all voxels in regions 55,56,70,71 (this can be seen in the second to left slide [-62] in region 70, where the top voxels in Figure \ref{fig:UKBsim_beta_true} within region 70 are 0). To ensure smoothness, we use this exponential square function to generate $\beta$, but this function is different from the GP kernel used in UKB analysis, where we used Mat\`ern kernel for each region and tuned the kernel parameters so that the estimated covariance can be similar to the empirical covariance of the observed data. The simulation study uses the same kernel as in the UKB analysis, without knowing any information about the smoothness in the true signal $\beta$ or using any fine-tuning according to the observed images.

The location of the four regions is highlighted in Figure \ref{fig:UKBsim_beta_region}. Based on Figures \ref{fig:UKBsim_beta_true} and \ref{fig:UKBsim_beta_region}, $\beta$ is generated with smoothly varying functional values across the region boundaries of region 55 and 56, and region boundaries of region 70 and 71. 

In addition, we compare the selection accuracy between MUA and SBIOSimp result. MUA's selection is determined by the BH-adjusted p-values less than a target threshold $\alpha$, whereas SBIOSimp's selection is made by thresholding PIP above $1-\alpha$. The result is shown in Table \ref{tb:UKBsim_selection}, which indicates that SBIOSimp can better control FDR. The higher false negatives in SBIOSimp result may be due to the GP kernel not being flexible enough, where small effects in the upper peripheral of region 70 tend to be neglected by SBIOSimp (as shown in the highlighted blue boxes in Figures \ref{fig:UKBsim_beta_true} and \ref{fig:UKBsim_beta_est}).

\begin{table}[]
\caption{Comparision between MUA and SBIOSimp selection on active $\beta(s)$, under varying target error $\alpha=0.1,0.05,0.01$.}
\label{tb:UKBsim_selection}
\resizebox{\columnwidth}{!}{
\begin{tabular}{lllllllll}
\toprule
\multicolumn{3}{c}{\textbf{Target $\alpha$ = 0.1}} & \multicolumn{3}{c}{\textbf{Target $\alpha$ = 0.05}} & \multicolumn{3}{c}{\textbf{Target $\alpha$ = 0.01}} \\
\multicolumn{3}{c}{MUA}              & \multicolumn{3}{c}{MUA}               & \multicolumn{3}{c}{MUA}              \\
         & est. FALSE   & est. TRUE  &          & est. FALSE   & est. TRUE   &          & est. FALSE   & est. TRUE  \\
true 0   & 111228       & 672        & true 0   & 111580       & 320         & true 0   & 111836       & 64         \\
true 1   & 0            & 9965       & true 1   & 0            & 9965        & true 1   & 0            & 9965       \\ 
\multicolumn{3}{c}{SBIOSimp}         & \multicolumn{3}{c}{SBIOSimp}          & \multicolumn{3}{c}{SBIOSimp}         \\
         & est. FALSE   & est. TRUE  &          & est. FALSE   & est. TRUE   &          & est. FALSE   & est. TRUE  \\
true 0   & 111898       & 2          & true 0   & 111899       & 1           & true 0   & 111900       & 0          \\
true 1   & 101          & 9864       & true 1   & 106          & 9859        & true 1   & 115          & 9850      \\
\bottomrule
\end{tabular}
}
\end{table}

\section{Low-resolution Simulation Study with a Single GP kernel}\label{sec:supp_oneKer}

The region-wise independence assumption provides not only computational convenience, it also allows us to capture more detailed structure within each region. However, this assumption is not required as long as the computational cost is manageable. We demonstrate the performance of SBIOSimp using a single GP kernel across a low-resolution brain image and compare with MUA using the following simulation example.

In this section, we design a low-resolution simulation study, where the entire fMRI data is contained in a cube of $45\times 54\times 45$ voxel space in $\R^3$, which is about $1/8$ of the original 2mm fMRI space of $91\times 109\times 91$ cube. In this low-resolution space, the computational constraint is relaxed, and we can perform the proposed SBIOSimp method with a single GP kernel. After preprocessing, we have $p=19091$ voxels to represent the whole brain mass, and will generate $n=10,000$ sample split into 10 batches.

\begin{figure}[ht!]
\centering
    \begin{subfigure}{\textwidth}
        \centering
        \includegraphics[width=\linewidth]{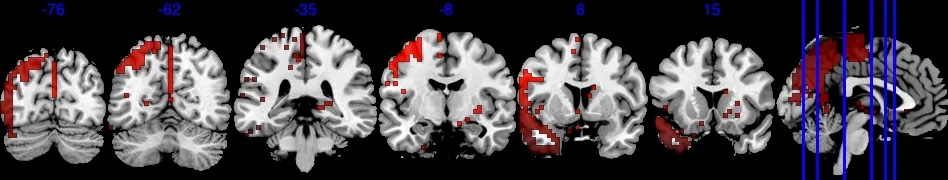}
        \caption{True $\beta(s)$ used in the low-resolution simulated data, where $\beta(s)$ is generated as a smooth function across 4 regions 55,56,70,71. Color ranges from back to bright red in [0.1,0.8].}
        \label{fig:lowd_beta_true}
    \end{subfigure}
    \begin{subfigure}{\textwidth}
        \centering
        \includegraphics[width=\linewidth]{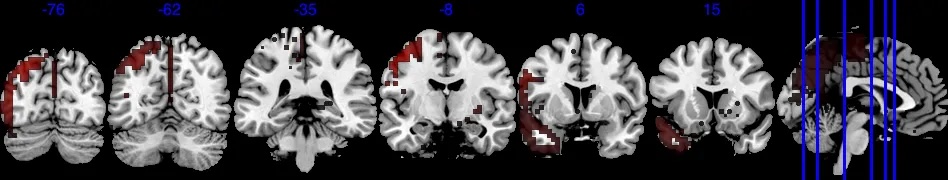}
        \caption{Estimated $\beta(s)$ by SBIOSimp, thresholded by PIP $\geq 0.9$. Color ranges from back to bright red in [0.1,0.8].}
        \label{fig:lowd_beta_est}
    \end{subfigure}
    \begin{subfigure}{\textwidth}
        \centering
        \includegraphics[width=\linewidth]{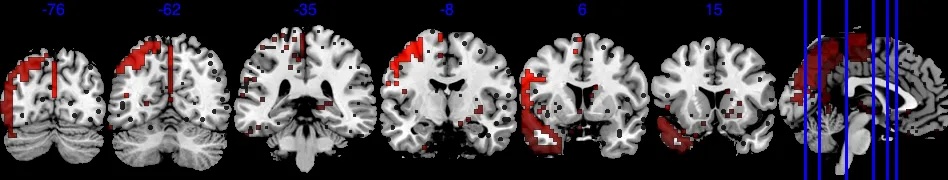}
        \caption{Estimated $\beta(s)$ by MUA, thresholded by the BH-adjusted p-value $\leq 0.1$.Color ranges from back to bright red in [0.1,0.8].}
        \label{fig:lowd_beta_MUA}
    \end{subfigure}
\caption{True, SBIOSimp and MUA estimated $\beta$ in the low-resolution simulated data.}
\label{fig:lowd_sim}
\end{figure}

First, we generate the true $\beta$ still using a smooth function $\beta(x) = \exp\{-0.01*\|x-x_0\|_2^2/d\}I_{x\in R_{55,56,70,71}}$, where $R_{55,56,70,71}$ is the collection of all voxels in the low-resolution brain, and $x_0$ is the geometric center of $R_{55,56,70,71}$. Hence $\beta$ is only piece-wise smooth across the whole brain. The true $\beta$ is shown in Figure \ref{fig:lowd_beta_true}. Note that Figure \ref{fig:lowd_sim} provides figures of low-resolution 3D images mapped into the original resolution space $91\times 109\times 91$, hence they look blurry and noisy.

The single GP kernel uses the Modified Squared Exponential Kernel
\[\kappa(x,x') = \exp\{-a(\|x\|_2^2 + \|x'\|_2^2) - b\|x-x'\|^2_2\}\]
and we set $a=0.01,b=1$, using 10 degrees of Hermite polynomials. This results in a total of $L=286$ basis. In addition, we apply QR-decomposition to the eigenvectors to make the basis functions approximately orthonormal. In this setting, the GP kernel may not well represent the true $\beta$ since they have different smoothness assumptions, and the true $\beta$ is only piecewise smooth but not a global smooth function over the whole brain mass.

Figure \ref{fig:lowd_sim} shows the estimation result for MUA and SBIOSimp respectively, and Table \ref{tb:lowd_sim} provides the selection accuracy for the two methods. Again, SBIOSimp tends to control FDR better than MUA, as illustrated both in Table \ref{tb:lowd_sim} and in the third and forth to the left slices in Figure \ref{fig:lowd_beta_est} and \ref{fig:lowd_beta_MUA}. Although due to the single GP kernel not exactly representing the smoothness of the true $\beta$, the scale of the estimated $\beta$ tends to be smaller than the truth. 

This example is also good for demonstrating the use of independent Bernoulli prior for $\delta$ parameter. Because we use an independent prior $\delta$ across the whole brain, there might be concerns that $\delta$ cannot capture the spatially varying pattern of the active area. This example shows that the independence is only on the prior
level, the posterior of $\delta(s)$ depends on $\beta(s)$, a spatially varying parameter structured by the corresponding GP kernel. On the other hand, the active region selected by MUA, a voxel-level analysis, completely ignores spatial correlation and thus gives very noisy results with a high FDR. 

The result in Figure \ref{fig:lowd_sim} also shows that even when the kernel is misspecified, SBIOSimp is still able to identify the active area, whereas the point estimation on the scale of $\beta$ is impacted by the misspecified kernel.

\begin{table}[ht!]
\caption{Low-resolution simulation with single GP kernel: Comparision between MUA and SBIOSimp selection on active $\beta(s)$, under varying target error $\alpha=0.1,0.05,0.01$.}
\label{tb:lowd_sim}
\resizebox{\columnwidth}{!}{
\begin{tabular}{lllllllll}
\toprule
\multicolumn{3}{c}{\textbf{Target $\alpha$ = 0.1}} & \multicolumn{3}{c}{\textbf{Target $\alpha$ = 0.05}} & \multicolumn{3}{c}{\textbf{Target $\alpha$ = 0.01}} \\
\multicolumn{3}{c}{MUA}              & \multicolumn{3}{c}{MUA}               & \multicolumn{3}{c}{MUA}              \\
         & est. FALSE   & est. TRUE  &          & est. FALSE   & est. TRUE   &          & est. FALSE   & est. TRUE  \\
true 0   & 17770    &137        & true 0   & 17835       & 72         & true 0   & 17895       & 12         \\
true 1   & 0            & 1184       & true 1   & 0            & 1184        & true 1   & 0            & 1184       \\ 
\multicolumn{3}{c}{SBIOSimp}         & \multicolumn{3}{c}{SBIOSimp}          & \multicolumn{3}{c}{SBIOSimp}         \\
         & est. FALSE   & est. TRUE  &          & est. FALSE   & est. TRUE   &          & est. FALSE   & est. TRUE  \\
true 0   & 17892       & 15          & true 0   & 17903       & 4           & true 0   & 17906       & 1          \\
true 1   & 9          & 1175       & true 1   & 12          & 1172        & true 1   & 12          & 1172      \\
\bottomrule
\end{tabular}
}
\end{table}

\end{document}